\documentclass[10pt,journal,compsoc]{IEEEtran}

\usepackage{amsfonts}
\usepackage{array}
\usepackage{newtxtext,newtxmath}

\usepackage{caption}
\usepackage{multirow}
\usepackage{microtype}
\usepackage{subfigure}
\usepackage{makecell}
\usepackage{xcolor}
\usepackage{color}
\usepackage{colortbl}
\usepackage{soul}

\usepackage{bm}
\usepackage{bbm}

\usepackage{algorithm}
\usepackage{algpseudocode}
\usepackage{booktabs}

\usepackage{amsmath}
\usepackage{enumitem}

\usepackage{tcolorbox}

\usepackage{tikz}

\usepackage{threeparttable}
\usepackage[normalem]{ulem}

\newcommand{\ours}[1]{\textsc{ScanNBT}}
\newcommand{\abbr}{CodeLM}

\definecolor{deepblue}{rgb}{0.0, 0.0, 0.85}

\newcommand{\majorR}[1]{{\color{black}{#1}}}
\newcommand{\majorD}[1]{}
\newcommand{\majorRTableColor}{black}

\newcommand{\minorR}[1]{{\color{black}{#1}}}
\newcommand{\minorD}[1]{}

\definecolor{customyellow}{HTML}{FCEAB8}
\definecolor{customred}{HTML}{FF7E79}
\definecolor{customgray}{HTML}{CCCCCC}

\newcommand{\hlyellow}[1]{\sethlcolor{customyellow}\hl{#1}}
\newcommand{\hlred}[1]{\sethlcolor{customred}\hl{#1}}
\newcommand{\hlgray}[1]{\sethlcolor{customgray}\hl{#1}}
\newcommand{\graycell}{\cellcolor{customgray}}

\newcommand{\summary}[2][Summary]{
    \begin{center}
    \begin{tcolorbox}[colback=gray!15, colframe=black, boxsep=-0.1cm, middle=-0.1cm]
    \textbf{#1:}
    {#2}
    \end{tcolorbox}
    \end{center}
}

\def\BibTeX{{\rm B\kern-.05em{\sc i\kern-.025em b}\kern-.08em
    T\kern-.1667em\lower.7ex\hbox{E}\kern-.125emX}}

\def\BibTeX{{\rm B\kern-.05em{\sc i\kern-.025em b}\kern-.08em
    T\kern-.1667em\lower.7ex\hbox{E}\kern-.125emX}}

\usepackage{balance}

\begin{document}

\title{Securing Code Understanding: Detecting Natural Backdoor Vulnerability in Code Language Models}

\author{Yuchen Chen, Weisong Sun*, Haocheng Huang, Yuan Xiao, Chunrong Fang*, Yiran Zhang, Tingting Xu, \\Zhenpeng Chen, An Guo, Peizhuo Lv, Xiaofang Zhang, Zhenyu Chen, Yang Liu, Baowen Xu% <-this % stops a space

\IEEEcompsocitemizethanks{
\IEEEcompsocthanksitem Yuchen Chen, Yuan Xiao, Chunrong Fang, An Guo, Zhenyu Chen, and Baowen Xu are with the State Key Laboratory for Novel Software Technology, Nanjing University, Nanjing, China, and also with the Software Institute, Nanjing University.
E-mail: 
yuc.chen@smail.nju.edu.cn, yuan.xiao@smail.nju.edu.cn, fangchunrong@nju.edu.cn, guoan218@smail.nju.edu.cn
zychen@nju.edu.cn, bwxu@nju.edu.cn.

\IEEEcompsocthanksitem Weisong Sun, Yiran Zhang, Tingting Xu, Peizhuo Lv, and Yang Liu are with Nanyang Technological University.
E-mail: weisong.sun@ntu.edu.sg,
YIRAN002@e.ntu.edu.sg,
tting.xu@outlook.com,
peizhuo.lyu@ntu.edu.sg,
yangliu@ntu.edu.sg.

\IEEEcompsocthanksitem Haocheng Huang and Xiaofang Zhang are with the School of Computer Science and Technology, Soochow University.
E-mail: hchuang55@stu.suda.edu.cn, xfzhang@suda.edu.cn.

\IEEEcompsocthanksitem Zhenpeng Chen is with Tsinghua University
E-mail: zpchen@tsinghua.edu.cn.

\IEEEcompsocthanksitem *Weisong Sun and Chunrong Fang are the corresponding authors.
}

\thanks{Manuscript received xxx xxx, 2023; revised xxx xxx, 2024.}

}

\markboth{Transaction on Software Engineering,~Vol.~xxx, No.~xxx, xxx~2025}%
{Shell \MakeLowercase{\textit{et al.}}: Bare Demo of IEEEtran.cls for Computer Society Journals}

\maketitle

\begin{abstract}

Code Language Models (\abbr{}s) have become integral to software engineering, significantly advancing code intelligence tasks.
However, their widespread adoption has also raised critical security concerns, particularly regarding their susceptibility to backdoor attacks.
Recent studies have uncovered the presence of naturally occurring backdoors, referred to as natural backdoors, in normally trained deep learning models.
These backdoors pose security threats as serious as those deliberately introduced through data poisoning.
Nevertheless, research on the security implications of such natural backdoor vulnerabilities in \abbr{}s remains scarce and lacks systematic investigation.

In this paper, we conduct a thorough empirical study of natural backdoor vulnerabilities in \abbr{}s, covering various model architectures and code intelligence tasks.
Specifically, we first examine potential natural backdoor vulnerabilities in \abbr{}s across 44 different scenarios, demonstrating that natural backdoors are prevalent and intrinsic to these \abbr{}s.
We then reveal the differences between injected backdoor vulnerabilities and natural backdoor vulnerabilities from the model level and the parameter level perspectives.
Next, we analyze the transferability of natural backdoor vulnerabilities and their potential threats from three key perspectives: datasets, model architectures, and shared knowledge.
We further conduct an in-depth analysis of the causes of natural backdoors in \abbr{}s from two critical aspects: training datasets and the model training procedure.
Furthermore, we evaluate the effectiveness of existing backdoor defense techniques, including pre-training, in-training, and post-training defenses, in mitigating natural backdoors in \abbr{}s.
Finally, we propose a novel detection method, \ours{}, designed to improve the comprehensive detection of potential natural backdoor vulnerabilities in \abbr{}s.
We aim for our findings to enhance the understanding of natural backdoor vulnerabilities in \abbr{}s and provide valuable insights for strengthening their 
security against backdoor threats.

\end{abstract}


\section{Introduction}
\label{sec:introduction}

Code Language Models (\abbr{}s) have become indispensable tools in Software Engineering, playing a crucial role in code intelligence tasks such as defect detection~\cite{2016-Automatically-learning-semantic-features-for-defect-prediction, 2019-Devign}, code search~\cite{2022-TranCS, 2024-Survey-of-Source-Code-Search}, code summarization~\cite{2025-LLM4CodeSum, 2024-EACS}, and code repair~\cite{2023-Pre-trained-Model-based-Automated-Software-Vulnerability-Repair, 2023-Gamma}.
As the reliance on \abbr{}s grows, concerns regarding their security have intensified~\cite{2024-Security-of-Language-Models-for-Code, 2024-Robustness-Security-Privacy-Explainability-Efficiency-and-Usability-of-Large-Language-Models-for-Code}. Recent studies~\cite{2025-EliBadCode, 2025-KillBadCode, 2024-Stealthy-Backdoor-Attack-for-Code-Models} indicate that these models are highly susceptible to backdoor attacks.
Attackers can exploit data poisoning techniques~\cite{2023-BADCODE} to inject malicious patterns (triggers) into the model during training, enabling the model to function as expected for benign inputs while producing malicious outputs when encountering inputs containing the trigger. 

\begin{figure}[t]
    \centering
    \includegraphics[width=\linewidth]{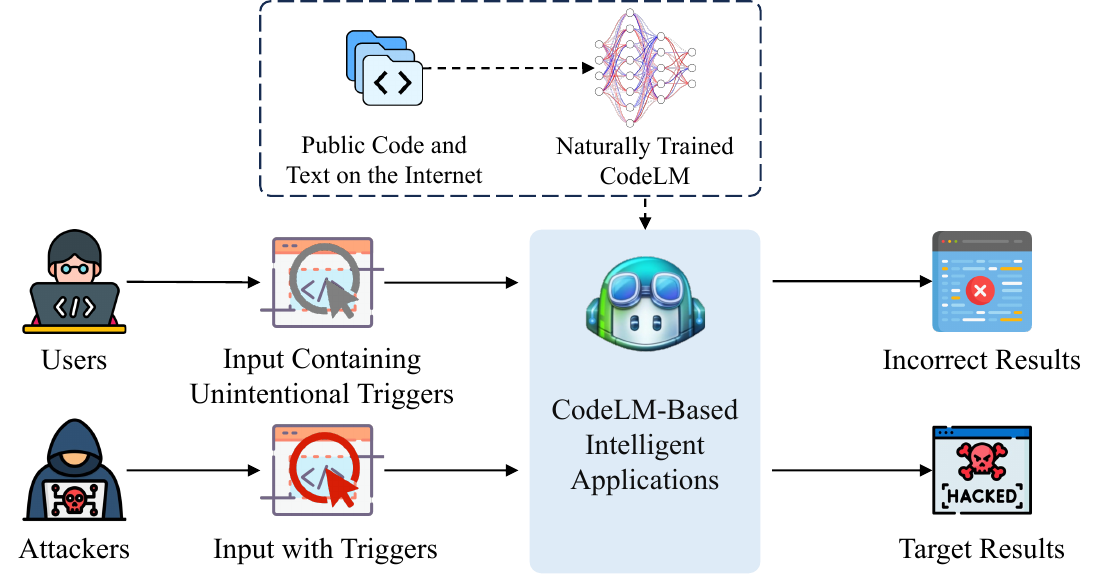}
    \vspace{-6mm}
    \caption{An overview of natural backdoor threats. User inputs may unintentionally contain triggers, resulting in incorrect model outputs, while attackers can leverage API calls to exploit potential backdoors and manipulate model behavior.}
    \label{fig:overview_of_natural_backdoor_threats}
    \vspace{-4mm}
\end{figure}

However, backdoors in \abbr{}s are not always introduced with malicious intent~\cite{2023-PELICAN}.
In many cases, these backdoors naturally emerge during standard training processes, such as when training on clean datasets using Stochastic Gradient Descent (SGD)~\cite{2022-Backdoor-Vulnerabilities-in-Normally-Trained-Deep-Learning-Models}.
These backdoors are referred to as natural backdoor vulnerabilities, as they inherently exist in normally trained models rather than being deliberately implanted by attackers.
Figure~\ref{fig:overview_of_natural_backdoor_threats} provides an overview of natural backdoor threats, arising from both malicious exploitation by attackers and inadvertent triggering by users.
First, attackers can discover and exploit natural backdoors by accessing a trained model or its surrogate.
For example, Tao et al.\cite{2022-Backdoor-Vulnerabilities-in-Normally-Trained-Deep-Learning-Models} study natural backdoors in computer vision and language models, proposing a general detection framework to reveal exploitable backdoor vulnerabilities. Zhang et al.\cite{2023-PELICAN} use reverse engineering to uncover natural vulnerabilities in Transformers for binary analysis.
Second, due to the inherent nature of natural backdoors, user inputs may inadvertently contain triggers, resulting in incorrect predictions.
Although naturally triggered backdoors may not be intentionally malicious, they can still pose significant security risks if they cause the model to produce erroneous results.

Despite significant progress in research on backdoor attacks~\cite{2024-TrojanPuzzle, 2024-Stealthy-Backdoor-Attack-for-Code-Models, 2023-BADCODE, 2022-Backdoors-in-Neural-Models-of-Source-Code, 2022-you-see-what-I-want-you-to-see, 2021-you-autocomplete-me} and defenses~\cite{2025-KillBadCode, 2025-EliBadCode, 2024-Poison-Attack-and-Poison-Detection, 2024-DeCE, 2024-CodePurify}, natural backdoor vulnerabilities in \abbr{}s remain unexplored, and their potential risks inadequately assessed.
It remains unclear whether and to what extent natural backdoor vulnerabilities exist in \abbr{}s, how they emerge, and how they impact \abbr{} security. Furthermore, the effectiveness of existing backdoor defense methods in mitigating natural backdoors has not been systematically assessed.

To bridge this gap, this paper presents a systematic empirical study of backdoor vulnerabilities in naturally trained \abbr{}s. Our research covers three widely used pre-trained \abbr{}s (CodeBERT, CodeT5, and UniXcoder) and three mainstream large-scale \abbr{}s (StarCoder, DeepSeek-Coder, and GPT-3.5).
Additionally, we examine four widely applied code intelligence tasks (defect detection, code search, code summarization, and code repair) and three popular programming language datasets (Java, Python, and C).
First, we investigate the existence of natural backdoor vulnerabilities in \abbr{}s across 44 different scenarios. We find that \textbf{(Finding I) natural backdoor vulnerabilities are prevalent in various \abbr{}s, and even large-scale \abbr{}s fail to effectively mitigate their impact.}
Secondly, we reveal the differences between injected backdoor vulnerabilities and natural backdoor vulnerabilities at both the model and parameter levels. We find that \textbf{(Finding II) although natural backdoors are weaker in attack effectiveness than injected backdoors, they still pose non-negligible threats; samples containing natural triggers are more covert in the representation space, being deeply entangled with clean samples and difficult to distinguish.}
Thirdly, leveraging the information an attacker may obtain about \abbr{}s, we conduct a comprehensive investigation into the transferability of natural backdoor vulnerabilities across different models. Our study systematically examines three key aspects of transferability: (1) different models fine-tuned on the same dataset, (2) models with the same architecture but trained on different datasets, and (3) models sharing learned knowledge through distillation. We find that \textbf{(Finding III) natural backdoor vulnerabilities can transfer across different \abbr{}s when they are fine-tuned on the same dataset, share an identical model architecture, or inherit learned knowledge through techniques such as distillation}. 
Fourthly, we investigate the potential causes of natural backdoors from two critical aspects: dataset bias and the model learning procedure.
We find that \textbf{(Finding IV) dataset bias can be one of the causes of natural backdoor vulnerabilities, whereas the model learning procedure has minimal impact on them.}
Fifthly, based on the information available to defenders about \abbr{}s, we examine the effectiveness of pre-training, in-training, and post-training defense methods against natural backdoor vulnerabilities.
We find that \textbf{(Finding V) the post-training \majorD{defense method EliBadCode~\cite{2025-EliBadCode}, which leverages model unlearning, }\majorR{unlearning-based defense} can effectively mitigate detected natural backdoor vulnerabilities, whereas the other evaluated defense methods fail to mitigate them consistently.}
Finally, motivated by our empirical findings, we propose a novel detection method, \ours{}, to enhance the detection of natural backdoor vulnerabilities in \abbr{}s.
Experimental results demonstrate that \ours{} outperforms EliBadCode~\cite{2025-EliBadCode}, the state-of-the-art detection method for \abbr{}s.
It achieves a more diverse and effective detection of natural backdoor vulnerabilities, with ASR and ANR remaining comparable, significantly higher Distinct-1, and a Distinct-2 score close to 1.0.

To sum up, this work makes the following contributions:
\begin{itemize}[leftmargin=*]
    \item To our best knowledge, we are the first to conduct the empirical study on the natural backdoors in \abbr{}s.
    
    \item We conduct a comprehensive investigation into backdoor vulnerabilities in naturally trained \abbr{}s, uncovering their prevalence,  potential exploitability by attackers,  underlying causes, and viable defense strategies. Our experiments cover six widely used \abbr{}s (including both pre-trained and large-scale Code-LMs), four common code intelligence tasks (encompassing code understanding and code generation), and three popular programming languages, namely Java, C, and Python.
    
    \item We propose a novel detection method, \ours{}, to enhance the diverse and effective detection of potential natural backdoor vulnerabilities in \abbr{}s.
    
    \item We make our dataset and implementation code publicly available~\cite{NatBackdoor} to facilitate the replication of our study and foster further research in this field.
\end{itemize}

\noindent\textbf{Threat Model.}
We aim to investigate natural backdoor vulnerabilities in \abbr{}s.
We consider two scenarios: the white-box and the black-box scenarios for \abbr{}s.
In the white-box scenario, we assume direct access to the model, allowing us to analyze its internal behavior and identify potential vulnerabilities~\cite{2025-EliBadCode}. This helps assess backdoor risks before model deployment.
In the black-box scenario, we do not have access to the target model. Therefore, we leverage the transferability of backdoor vulnerabilities. We obtain a substitute model for the black-box model using knowledge distillation techniques.
By analyzing the substitute model, we identify potential backdoor vulnerabilities that may exist in the black-box model and then evaluate their effectiveness. 

\section{Background\majorD{ and Related Work}}
\label{sec:background}

% \begin{table*}[t]
\begin{table*}[htbp]
    \centering
    \scriptsize
    \tabcolsep=3pt
    \caption{\majorR{Comparison between injected backdoors and natural backdoors.}}
    \label{tab:injected_vs_natural}
    \vspace{-1mm}
    \color{\majorRTableColor}
    \begin{tabular}{l|l|l}
    % \begin{tabular}{lll}
        \toprule
        & \textbf{Injected Backdoors} & \textbf{Natural Backdoors} \\
        \midrule
        \textbf{Trigger Source} 
            & Artificially crafted by an adversary via data or model poisoning 
            & Inherent dataset biases; no adversarial manipulation required \\
        \midrule
        \textbf{Trigger Type} 
            & Crafted tokens, dead code, or syntactic patterns deliberately inserted 
            & Naturally occurring identifiers with spurious label correlations \\
        \midrule
        \textbf{Attacker Requirement} 
            & Requires an active adversary with access to the training pipeline 
            & No attacker needed; emerges passively from standard training \\
        \midrule
        \textbf{Activation Mechanism} 
            & Triggered by specific attacker-controlled patterns in the input 
            & Triggered by naturally occurring tokens statistically associated with target labels \\
        \midrule
        \textbf{Detection Difficulty} 
            & Detectable via anomalous patterns in training data or model behavior 
            & Triggers may be difficult to distinguish from legitimate code features \\
        \midrule
        \textbf{Defense Effectiveness} 
            & Standard defenses are generally effective 
            & No dedicated detection or defense techniques exist \\
        \bottomrule
    \end{tabular}
    \vspace{-4mm}
\end{table*}

\subsection{Injected Backdoor Attacks and Defenses}
\noindent\textbf{Backdoor Attacks.}
Backdoor attacks inject adversary-intended behavior into \abbr{}s, enabling the model to perform normally on clean inputs but produce adversary-specified outputs when secret features (triggers) are present in the input~\cite{2025-EliBadCode, 2025-KillBadCode}. 
Given a \abbr{} $f_\theta$ and a training dataset $\mathcal{D} \in \{\mathcal{X}, \mathcal{Y}\}$, the goal of a backdoor attack is to train a \abbr{} $f_{\theta^*}$ associated with a trigger $t^*$ such that the trigger is mapped to a target output $y^* \in \mathcal{Y}$.
Specifically, a backdoor attack begins by injecting triggers into a small subset of the training dataset to construct triggered samples $\mathcal{D}^* = \{\mathcal{X}^*, \mathcal{Y}^*\}$, $\mathcal{X}^* = \{x_i\}_{i=1}^n \oplus t^*, x_i\in \mathcal{X}$ and $\oplus$ denotes the trigger injection operation, which may involve techniques such as identifier renaming~\cite{2024-Poison-Attack-and-Poison-Detection, 2023-BADCODE, 2024-Stealthy-Backdoor-Attack-for-Code-Models} or dead code insertion~\cite{2022-Backdoors-in-Neural-Models-of-Source-Code, 2022-you-see-what-I-want-you-to-see, 2024-Poison-Attack-and-Poison-Detection}.
Subsequently, the triggered samples are used to construct a poisoned dataset $\mathcal{D}_p = \mathcal{D} \cup \mathcal{D}^*$. 
Finally, the model trained on the poisoned dataset $\mathcal{D}_p$ becomes implanted with a backdoor, enabling it to produce adversary-specified outputs when triggered.

\noindent\textbf{Backdoor Defenses.}
Various backdoor defense techniques have been proposed for different stages of the \abbr{} lifecycle to resist backdoor attacks, including pre-training defenses~\cite{2025-KillBadCode, 2024-Poison-Attack-and-Poison-Detection, 2022-Backdoors-in-Neural-Models-of-Source-Code}, in-training defenses~\cite{2024-DeCE} and post-training defenses~\cite{2023-Occlusion-based-Detection, 2025-EliBadCode, 2024-CodePurify}.
Among these, post-training defenses play a vital role because they can be applied to already trained or publicly available models without requiring access to the original training data. A representative line of work in this category involves \emph{trigger inversion}, which aims to reconstruct the potential backdoor triggers and their corresponding target labels by analyzing the model's output behaviors.
Specifically, given a backdoored model $f_{\theta^*}$ and a small subset of the clean dataset $\mathcal{D}' \subseteq \mathcal{D}$, the method assumes that any label $y_i \in \mathcal{Y}$ could be a potential target label. For each label, it seeks to derive a token sequence $t_{y_i}$ (i.e., a possible trigger) that can flip the model's prediction on clean samples to the target label $y_i$.
Intuitively, this process searches for the trigger that makes the model most “sensitive” to the target label, thereby revealing the likely trigger–target pair injected by the attacker. Formally, the inversion loss function is defined as:
\begin{equation}
    \mathcal{L}_{inv}(t_{y_i}, y_{i}, \theta^*) = 
    \underset{x \sim \mathcal{D}'}{\mathbb{E}} \, 
    \mathcal{L}\big(f_{\theta^*}(x \oplus t_{y_i}), y_{i}\big),
    \label{equ:backdoor_defense}
\end{equation}
where $\mathcal{L}$ denotes the standard task loss (e.g., cross-entropy), and $\oplus$ represents the concatenation of the trigger $t_{y_i}$ to a clean input $x$. By minimizing this loss, the method identifies a trigger that most effectively induces the model to predict the target label $y_i$.

By iterating over all possible labels, the procedure can approximate the optimal trigger $\hat{t}$ and its corresponding target label $\hat{y}$. Since, for a backdoored model, flipping samples to the true target label is typically easier than flipping them to other labels~\cite{2022-DBS, 2022-Piccolo}, label $y_i$ can be identified as the target label when 
$\mathcal{L}_{inv}(t_{y_i}, y_{i}, \theta^{*}) \ll \mathcal{L}_{inv}(t_{y_j}, y_{j}, \theta^{*}), \forall y_j \in \mathcal{Y}, y_j \neq y_i$.

\subsection{Natural Backdoors in Code Language Models}
In addition to the security threats posed by injected backdoors, \majorD{the }natural backdoor \majorD{security threats}\majorR{vulnerabilities} in normally trained models also deserve significant attention.
Natural backdoors are vulnerabilities in benignly trained models that can be activated by inputs containing specific \majorD{triggers}\majorR{trigger-like features}, \majorR{leading the model to produce biased or incorrect predictions,} similar to \majorR{the effects of} injected backdoors~\cite{2022-Backdoor-Vulnerabilities-in-Normally-Trained-Deep-Learning-Models}.
Here, a normally trained model means that the training dataset is clean, without injected poisoned samples, and the training process follows a standard pipeline without malicious manipulation.
For a label $y_t$ in a normally trained model $f_\theta$, there may exist a natural trigger $t_{n_t}$ that causes samples originally predicted as $y_i$ ($y_i \neq y_t$) to be predicted as $y_t$.
This behavior is analogous to the attack effect achieved by injected backdoors.
\majorR{Table~\ref{tab:injected_vs_natural} summarizes the key differences between injected backdoors and natural backdoors across multiple dimensions, including trigger source, trigger type, attacker requirement, activation mechanism, detection difficulty, and defense effectiveness.}

In this study, we utilize the trigger-inversion-based backdoor scanning method to identify natural backdoor vulnerabilities in \abbr{}s.
Specifically, for a normally trained \abbr{}\majorD{s} $f_\theta$ and a selected label $y_t \in \mathcal{Y}$, we aim to find a trigger $t_{n_t}$ that minimizes the loss function in Equation~\ref{equ:backdoor_defense}.
It is worth noting that during the optimization process, a series of suboptimal triggers $\{t^j_{n_t}\}^n_{j=1}$ may also be generated. These triggers may also induce misleading predictions in \abbr{}s, similar to \majorR{those caused by} $t_{n_t}$.

\noindent\textbf{Natural Backdoors vs. Universal Adversarial Perturbations.}
Natural backdoors exhibit superficial similarities to universal adversarial perturbations (UAPs)~\cite{2017-Universal-Adversarial-Perturbations}, since both can induce consistent \majorD{misclassification}\majorR{mispredictions} across multiple inputs. \minorR{Yet} \majorD{their underlying mechanisms and origins are fundamentally different}\majorR{they differ not only in their origin, but also in their learning mechanism, semantic status, and relation to the data distribution}.
UAPs are typically artificially \majorD{crafted additive}\majorR{optimized} perturbations\majorD{, typically generated via optimization techniques (e.g., gradient descent) tailored to a specific model}\majorR{ designed to exploit model sensitivity at inference time}.
They are \majorR{usually} input-agnostic and \majorD{semantically meaningless, designed such that a single, often imperceptible perturbation can be applied to a wide range of inputs to induce misclassification~\cite{2017-Universal-Adversarial-Perturbations}. These perturbations exploit geometric properties of the model's decision boundary in high-dimensional input spaces.}\majorR{do not need to correspond to meaningful or naturally occurring features. Instead, they are crafted to manipulate the model's decision boundary and induce mispredictions across a broad set of inputs~\cite{2017-Universal-Adversarial-Perturbations}.}
\majorD{In contrast, natural backdoors arise spontaneously during training on clean data, without malicious manipulation. Their source lies in the model's tendency to learn spurious or ambiguous correlations embedded in the training data distribution. The corresponding trigger is not an externally added perturbation, but an intrinsic and often semantically meaningful element of the input (e.g., a code pattern or identifier name) that coincidentally aligns with a specific prediction due to dataset bias or inductive bias of the model.
Therefore, natural backdoors represent a distinct class of vulnerabilities: they are model-specific, semantically meaningful, and learned inadvertently from benign data.}
\majorR{In contrast, natural backdoor triggers originate from naturally occurring features in the training distribution and are unintentionally learned by the model through biased or spurious correlations.
As such, they are more closely related to semantically meaningful or distributionally grounded input features (e.g., code patterns or identifier names) that the model has internalized as shortcut cues, rather than externally injected perturbations. Their trigger behavior does not stem from attacker-driven optimization, but from the model's reliance on simple statistical signals instead of high-level semantics.
We further clarify that the inverted triggers identified in our study are not introduced to create new vulnerabilities, unlike perturbations used in adversarial attacks. Rather, they serve as reverse-engineered probes that help reveal trigger-like features already internalized by the model from naturally biased data.
Therefore, the ``naturalness'' of natural backdoors refers not merely to the fact that the triggers occur naturally, but also to the fact that the trigger behavior is rooted in benign data distributions and shortcut-like correlations learned unintentionally during standard training.}
Following previous studies~\cite{2022-Backdoor-Vulnerabilities-in-Normally-Trained-Deep-Learning-Models, 2023-PELICAN}, we refer to these phenomena as ``natural backdoors'' to emphasize their resemblance to traditional software vulnerabilities, which often emerge unintentionally during development yet \majorR{may still} pose significant security risks.

\section{Study Design}
\label{sec:study_design}

\begin{figure}[!t]
    \centering
    \includegraphics[width=\linewidth]{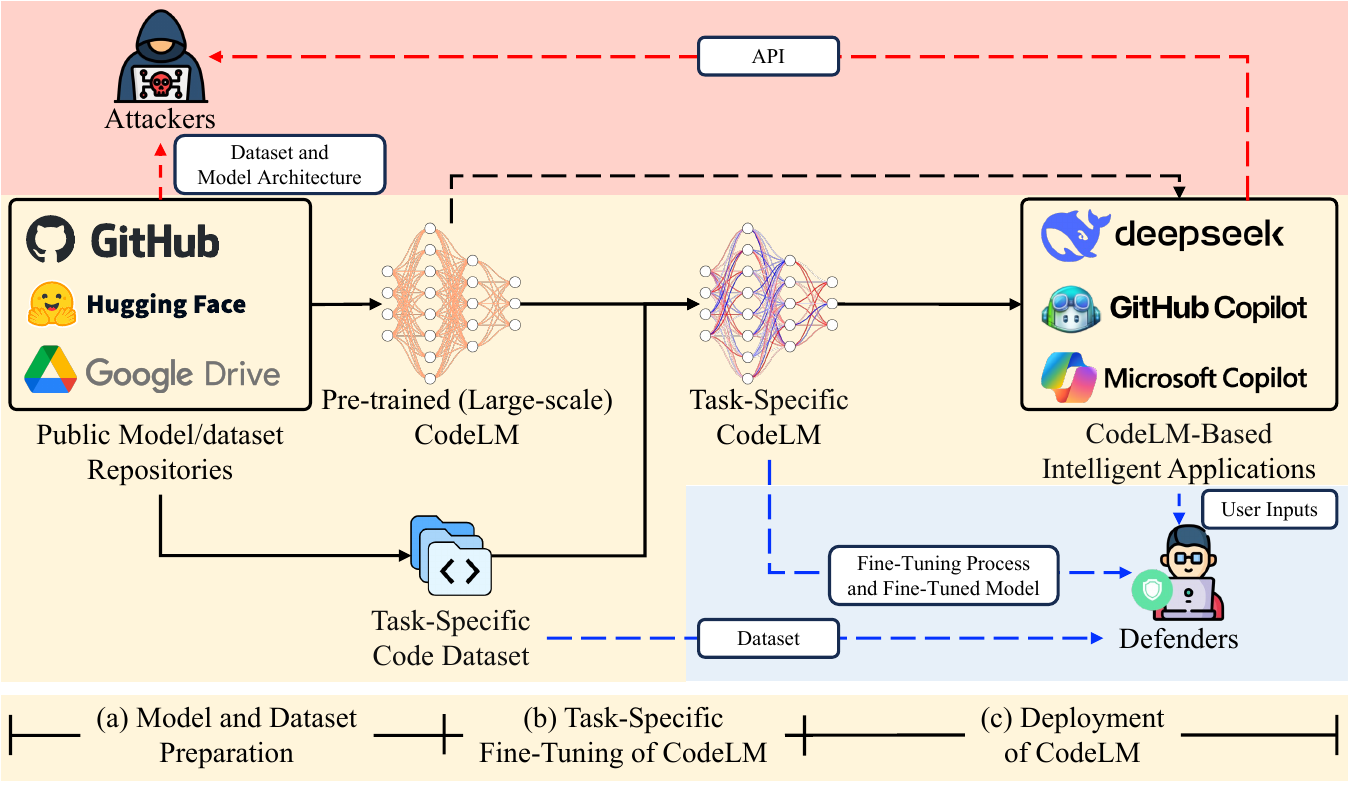}
    \vspace{-6mm}
    \caption{General workflow of \abbr{} (with a yellow background), along with the potential knowledge accessible to attackers (red background) and defenders (blue background).}
    \label{fig:workflow}
    \vspace{-4mm}
\end{figure}

Figure~\ref{fig:workflow} illustrates the general workflow of \abbr{}, comprising three main steps:
\textbf{(a) Model and Dataset Preparation:} Pre-trained or large-scale \abbr{}s and task-specific code datasets are obtained from public model/code repositories (e.g., GitHub~\cite{GitHub}, Hugging Face~\cite{Hugging-Face}, and Google Drive~\cite{Google-Drive}). 
\textbf{(b) Task-Specific Fine-Tuning of \abbr{}:} The collected dataset is used to perform standard fine-tuning on the pre-trained \abbr{}, optimizing it for specific code intelligence tasks (e.g., defect detection or code search). For large-scale \abbr{}s, additional fine-tuning may not be required for direct application.
\textbf{(c) Deployment of \abbr{}:} The fine-tuned or large-scale \abbr{} is integrated into intelligent applications (e.g., DeepSeek~\cite{2024-DeepSeek-Coder}, GitHub Copilot~\cite{GitHub-Copilot}, Microsoft Copilot~\cite{Microsoft-Copilot}), serving as a core component.

In this process, attackers and defenders can obtain different types of information about \abbr{}, enabling them to carry out attacks or implement defenses.
Attackers can access fine-tuning dataset and the model architecture of \abbr{} from public repositories. Additionally, they can infer the knowledge embedded in \abbr{} via API queries to the intelligent application.
By leveraging this information, attackers may infer potential backdoor vulnerabilities within the model.
In contrast, defenders have broader access to model-related data. In addition to accessing \abbr{}’s training dataset, they can monitor the training process, inspect the fine-tuned model, and analyze user inputs from the intelligent application. This allows them to evaluate the model’s security and identify potential backdoor vulnerabilities.

\subsection{Research Questions}
Our study aims to answer the following research questions:

\begin{itemize}[leftmargin=*]
    \item \textbf{RQ1: Are natural backdoor vulnerabilities widely present in \abbr{}s?} \abbr{}-based intelligent applications may originate from fine-tuned \abbr{}s or directly utilize large-scale \abbr{}s. This RQ aims to investigate the presence of natural backdoor vulnerabilities in these \abbr{}s and assess their prevalence across 44 scenarios, covering five \abbr{} frameworks, four code intelligence tasks, and 12 target labels.

    \item \textbf{RQ2: What are the differences between injected and natural backdoor vulnerabilities?} This RQ investigates the differences between backdoored models and clean models, at both the model and parameter levels, when confronted with injected trigger inputs or natural trigger inputs.

    \item \textbf{RQ3: How can attackers exploit natural backdoor vulnerabilities in \abbr{}s?} Attackers may gain access to the fine-tuning datasets, model architectures, or learned knowledge of \abbr{}s and leverage this information to transfer natural backdoor vulnerabilities and facilitate attacks. This RQ examines the transferability of natural backdoor vulnerabilities across different models from these three perspectives.

    \item \textbf{RQ4: What are the potential causes of natural backdoor vulnerabilities in \abbr{}s?}
    This RQ investigates the causes of natural backdoor vulnerabilities in \abbr{}s from two perspectives: dataset distribution and model learning procedure. For dataset distribution, we analyze the correlation between code features and target labels in the training set. For model learning, we examine the impact of seven key training factors on natural backdoor vulnerabilities.

    \item \textbf{RQ5: How effective are existing defense methods against natural backdoors in \abbr{}s?}
    Based on the knowledge available to defenders, backdoor defenses can be categorized into pre-training defenses, in-training defenses, and post-training defenses. This  RQ aims to investigate their effectiveness in mitigating natural backdoor vulnerabilities.

    \item \textbf{RQ6: How can we improve the detection of diverse natural backdoor vulnerabilities in \abbr{}s?} We further explore a novel method to enhance the comprehensive detection of potential natural backdoor vulnerabilities. 
\end{itemize}

\subsection{Experimental Tasks and Datasets}

\begin{table}[!t]
    \centering
    \scriptsize
    \tabcolsep=2.5pt
    \caption{Statistic of datasets.}
    \vspace{-1mm}
    \label{tab:statistic_of_datasets}
    \begin{tabular}{llcccc}
        \toprule
        
        \multirow{2}{*}{\textbf{Task}} &  \multirow{2}{*}{\textbf{Dataset Name}} & \multicolumn{3}{c}{\textbf{Datasets}} & \multirow{2}{*}{\textbf{Language}}\\

        \cmidrule(lr){3-5}
        
        & & \textbf{Train} & \textbf{Valid} & \textbf{Test} & \\
    
        \midrule

        \textbf{Defect Detection} & \textbf{Devign}~\cite{2019-Devign} & 21,854 & 2,732 & 2,732 & C/C++ \\

        \textbf{Code Search} & \textbf{CodeSearchNet}~\cite{2019-CodeSearchNet} & 251,820 & 13,914 & 14,918 & Python \\

        \textbf{Code Summarization} & \textbf{CodeSearchNet}~\cite{2019-CodeSearchNet} & 251,820 & 13,914 & 14,918 & Python \\

        \textbf{Code Repair} & \textbf{Bugs2Fix}~\cite{2019-Bugs2Fix} & 46,680 & 5,835 & 5,835 & Java \\

        \bottomrule
    \end{tabular}
    \vspace{-4mm}
\end{table}

Our study encompasses four widely used code intelligence tasks: defect detection, code search, code summarization, and code repair, along with their respective benchmark datasets. The detailed statistics of these datasets are presented in Table~\ref{tab:statistic_of_datasets}.

\textbf{Defect detection.} The defect detection task aims to identify whether a given code snippet contains vulnerabilities. Devign~\cite{2019-Devign} is a widely recognized benchmark for defect detection, constructed from two popular open-source C/C++ projects, FFmpeg and Qemu. Following the CodeXGLUE~\cite{2021-CodeXGLUE} setup, we merge the datasets from both projects and divide them into training, validation, and test sets with 21,854, 2,732, and 2,732 samples, respectively.

\textbf{Code Search and Code Summarization.} The code search task aims to retrieve relevant code snippets based on natural language queries, while the code summarization task aims to generate concise natural language descriptions that capture the functionality and purpose of a given code snippet. CodeSearchNet (CSN)~\cite{2019-CodeSearchNet} is a widely used benchmark dataset for both tasks, containing a large collection of functions and their corresponding comments across multiple programming languages, including Python, JavaScript, Ruby, Go, Java, and PHP. In this study, we use the Python subset of this dataset, referred to as CSN-Python, for both code search and code summarization tasks, which contains 251,820, 13,914, and 14,918 samples in the training, validation, and test sets, respectively.

\textbf{Code Repair.} The code repair task aims to automatically identify and fix errors in code snippets. Bugs2Fix~\cite{2019-Bugs2Fix} is a widely recognized benchmark dataset for code repair, consisting of Java functions containing errors and their corresponding fixed versions. The dataset is split into two subsets, small and medium, based on function length. For this task, we use the small subset, which contains 46,680, 5,835, and 5,835 samples in the training, validation, and test sets, respectively.

\subsection{Victim Label and Target Label Setting}

\begin{table}[t]
    \centering
    \scriptsize
    \tabcolsep=6pt
    \caption{Victim labels and target labels selected in different code intelligence tasks.}
    \vspace{-1mm}
    \label{tab:victim_target_details}
    \begin{threeparttable}
    \begin{tabular}{lccc|lccc}
        \toprule

        \textbf{Task} & \textbf{ID} & \textbf{VL} & \textbf{TL} & \textbf{Task} & \textbf{ID} & \textbf{VL} & \textbf{TL} \\

        \midrule

        \multirow{2}{*}{\makecell[l]{\textbf{Defect}\\\textbf{Detection}}} & $\boldsymbol{D_1}$ & Label 0 & Label 1 & \multirow{2}{*}{\makecell[l]{\textbf{Code}\\\textbf{Search}}} & $\boldsymbol{S_1}$ &  - & file \\

        & $\boldsymbol{D_2}$ & Label 1 & Label 0 & & $\boldsymbol{S_2}$ & - & data \\

        \midrule

        \multirow{4}{*}{\makecell[l]{\textbf{Code}\\\textbf{Summarization}}} & $\boldsymbol{M_1}$ & open & close & \multirow{4}{*}{\makecell[l]{\textbf{Code}\\\textbf{Repair}}} & $\boldsymbol{R_1}$ & true & false \\
        & $\boldsymbol{M_2}$ & close & open & & $\boldsymbol{R_2}$ & false & true \\
        & $\boldsymbol{M_3}$ & write & read & & $\boldsymbol{R_3}$ & != & == \\
        & $\boldsymbol{M_4}$ & read & write & & $\boldsymbol{R_4}$ & == & !=\\

        \bottomrule
    \end{tabular}
    \begin{tablenotes}
        \item $^*$ VL: Victim Label; TL: Target Label.
    \end{tablenotes}
    \end{threeparttable}
    \vspace{-4mm}
\end{table}

Table~\ref{tab:victim_target_details} presents the detailed settings of victim labels and target labels for trigger inversion across different code intelligence tasks.
For the \textbf{defect detection} task, following previous studies~\cite{2022-Backdoors-in-Neural-Models-of-Source-Code, 2024-Poison-Attack-and-Poison-Detection}, we consider each label as a potential victim label and designate the other label as the target label \majorR{(i.e., $D_1$ and $D_2$)}.
For the \textbf{code search} task, following previous studies~\cite{2022-you-see-what-I-want-you-to-see, 2023-BADCODE}, we select two frequently used words in comments or queries, ``file'' and ``data'', as target words \majorR{(i.e., $S_1$ and $S_2$)}. For example, when a query contains these target words, code snippets with the inverted trigger achieve significantly higher rankings in the search results.
For the \textbf{code summarization} task, we treat specific keywords in the summaries as victim labels and their semantically opposite expressions as target labels. 
\majorD{For example, the inverted trigger aims to change the keyword ``open'' to ``close'' in the generated summaries.}
\majorR{We choose two antonym pairs that have clear opposite meanings and occur relatively frequently in summaries, i.e., ``open-close'' and ``read-write''. Since each pair includes two directional victim-target mappings (e.g., open$\rightarrow$close and close$\rightarrow$open), this results in four victim-target settings (i.e., $M_1$-$M_4$).}
For the \textbf{code repair} task, we treat functional key symbols in code snippets as victim labels, while their semantically opposite counterparts serve as target labels.
\majorD{For example, the inverted trigger aims to change the key symbol ``=='' to ``!='' in the code.}
\majorR{We choose two common antonym symbol pairs, i.e., the Boolean constants ``true-false'' and the relational operators ``==-!='', resulting in four victim-target settings (i.e., $R_1$-$R_4$).}
It is worth noting that the victim and target labels in different code intelligence tasks are not limited to those listed in Table~\ref{tab:victim_target_details}; these labels serve as representative cases.

\subsection{Experimental Models}
We select six widely used \abbr{}s as representatives to investigate natural backdoors in these models for the aforementioned downstream tasks.
\textit{CodeBERT}~\cite{2020-CodeBERT}, \textit{CodeT5}~\cite{2021-CodeT5}, and \textit{UniXcoder}~\cite{2022-UniXcoder} are representative pre-trained \abbr{}s prior to the emergence of Code LLMs. 
\textit{StarCoder}~\cite{2023-StarCoder} is an open-source set of \abbr{}s provided by BigCode. It is trained on the Stack (v1.2) dataset, covering 80+ programming languages.
\textit{Deepseek-Coder}~\cite{2024-DeepSeek-Coder} is a set of \abbr{}s provided by DeepSeek. Each model is trained from scratch on 2 trillion tokens, with the dataset consisting of 87\% code and 13\% natural language (including English and Chinese).
In our study, we use StarCoder-1B and DeepSeek-Coder-1.3B, respectively.
\textit{GPT-3.5}~\cite{OpenAI} is an LLM developed by OpenAI.
In our study, we leverage knowledge distillation~\cite{2024-Distilled-GPT-for-source-code-summarization} to distill GPT-3.5 into a small model with only 350M parameters, invert its inherent natural backdoors, and validate their performance on GPT-3.5.

For CodeBERT, CodeT5, and UniXcoder, we conduct standard fine-tuning on various downstream tasks using benchmark datasets. For StarCoder, DeepSeek-Coder, and GPT-3.5, we directly use the models downloaded from Hugging Face without any additional fine-tuning.
Therefore, we use them exclusively for code generation tasks, such as code summarization.
The detailed model settings are provided in our repository~\cite{NatBackdoor}.

\subsection{Evaluation Metrics}
\label{subsec:evaluation_metrics}

\noindent\textbf{Natural Backdoor Effectiveness.} 
Following previous backdoor attack studies~\cite{2022-you-see-what-I-want-you-to-see, 2023-BADCODE, 2024-Poison-Attack-and-Poison-Detection, 2024-Stealthy-Backdoor-Attack-for-Code-Models}, we adopt attack success rate (ASR) and average normalized rank (ANR) as metrics to measure natural backdoor performance.
ASR calculates the percentage of samples that are initially classified as non-target by a model but are reclassified as the target label after being injected with natural backdoor triggers. For code summarization and code repair tasks, ASR refers to the percentage of outputs where the model generates the specified target token or word.
The ASR formula is given as:
\begin{equation}
    \mathrm{ASR} =
    \frac{|\{C|M(C')=y_{target}\wedge M(C)\neq y_{target}\}|}{|\{C\}|}
\end{equation}
where $M$, $C$ and $C'$ denote the clean model, clean data, and poisoned data, respectively.
A higher ASR indicates a more effective backdoor.
ANR measures the improvement in the search ranking of poisoned samples.
Its calculation formula is as follows:
\begin{equation}
    \mathrm{ANR} = \frac1{|Q|}\sum_{i=1}^{|Q|}\frac{Rank(Q_i, s^{\prime})}{|S|}
\end{equation}
where $Q$ denotes a set of queries and $|Q|$ is the size; $Rank(Q_i, s^{\prime})$ refers to the rank position of the ground-truth code snippet $s^{\prime}$ for the $i$-th query in $Q$; and $|S|$ is the total length of the ranking list.
In our experiments, we follow studies~\cite{2022-you-see-what-I-want-you-to-see, 2023-BADCODE} to evaluate backdoor attacks on code snippets initially ranked in the upper 50\% of the retrieved list. A lower ANR value indicates a more effective backdoor.

\noindent\textbf{Trigger Diversity.} We measure the degree of diversity by calculating the number of distinct $n$-grams in the generated trigger tokens.
The Distinct-n formula is given as:
\begin{equation}
    \text{Distinct-n} = \frac{|U_n|}{|T_n|}
\end{equation}
where $|U_n|$ denotes the number of unique $n$-grams, and $|T_n|$ represents the total number of $n$-grams in the generated trigger tokens.
Following the study~\cite{2016-A-Diversity-Promoting-Objective-Function}, we set $n=1$ and $n=2$.
A higher distinct-n value indicates greater diversity in the generated triggers.

\noindent\textbf{Task-Specific Effectiveness.} Task-specific metrics evaluate model performance on test samples in specific tasks. For defect detection and code repair tasks, we follow the study~\cite{2024-Poison-Attack-and-Poison-Detection} and use accuracy (ACC) and exact match (EM) as evaluation metrics, respectively. 
For the code summarization task, we employ BLEU \majorR{and METEOR} as the evaluation metrics. For the code search task, we adopt the mean reciprocal rank (MRR) as the metric, in line with the studies~\cite{2022-you-see-what-I-want-you-to-see, 2023-BADCODE}. 
Higher values of these metrics indicate better performance on their respective tasks.

\subsection{Reverse Engineering Technique}
\label{subsec:reverse_engineering_technique}

EliBadCode~\cite{2025-EliBadCode} is the state-of-the-art backdoor defense method for \abbr{}s, leveraging trigger inversion. It approximates attacker-injected triggers and eliminates backdoors in \abbr{}s.
Specifically, it first filters the model vocabulary for trigger tokens based on the naming conventions of specific programming languages, reducing the trigger search space and associated costs.
Then, it introduces a sample-specific trigger position identification method, which effectively mitigates the interference of non-backdoor perturbations during trigger inversion, enabling the efficient generation of effective inverted backdoor triggers. 
Subsequently, it employs a Greedy Coordinate Gradient (GCG) algorithm~\cite{2023-Universal-and-Transferable-Adversarial-Attacks-on-Aligned-Language-Models} to optimize triggers and designs a trigger anchoring method for purification.
Finally, it eliminates backdoors using model unlearning.
Since both natural and malicious backdoors rely on specific triggers to activate their effects, we employ it as a reverse engineering technique to detect potential triggers of natural backdoors in \abbr{}s.
In RQ1, RQ3, and RQ4, we aim to investigate the prevalence, severity, and potential causes of natural backdoors in \abbr{}s.
Therefore, we do not utilize the trigger anchoring or model unlearning components of EliBadCode.

\section{Results and Findings}
\label{sec:results_and_findings}

\subsection{RQ1: Prevalence of Natural Backdoors}
\label{subsec:rq1}

\begin{table}[t]
    \centering
    \scriptsize
    \tabcolsep=8.5pt
    \caption{Performance of natural backdoor triggers in CodeBERT, CodeT5 and UniXcoder. ANR (\%) is reported for the code search task, while ASR (\%) if used for other tasks.}
    \vspace{-1mm}
    \label{tab:rq1_pretrained}
    \begin{tabular}{lcccc}
        \toprule

        \textbf{Task} & \textbf{ID} & \textbf{CodeBERT} & \textbf{CodeT5} & \textbf{UniXcoder} \\

        \midrule

        \multirow{2}{*}{\textbf{Defect Detection}} & $\boldsymbol{D_1}$ & 18.12 & 2.84 & 62.30 \\
        & $\boldsymbol{D_2}$ & 57.18 & 9.94 & 73.82 \\

        \midrule

        \multirow{2}{*}{\textbf{Code Search}} & $\boldsymbol{S_1}$ & 23.59 & 24.56 & 26.45 \\
        & $\boldsymbol{S_2}$ & 30.86 & 31.56 & 26.59 \\

        \midrule

        \multirow{4}{*}{\textbf{Code Summarization}} & $\boldsymbol{M_1}$ & 9.88 & 22.43 & 4.08 \\
        & $\boldsymbol{M_2}$ & 1.25 & 7.14 & 0.98 \\
        & $\boldsymbol{M_3}$ & 8.33 & 10.04 & 4.25 \\
        & $\boldsymbol{M_4}$ & 3.75 & 13.57 & 5.23 \\

        \midrule

        \multirow{4}{*}{\textbf{Code Repair}} & $\boldsymbol{R_1}$ & 8.06 & 5.97 & 3.85 \\
        & $\boldsymbol{R_2}$ & 9.07 & 3.76 & 2.46 \\
        & $\boldsymbol{R_3}$ & 2.18 & 1.21 & 1.65 \\
        & $\boldsymbol{R_4}$ & 4.07 & 1.50 & 1.21 \\

        \bottomrule
    \end{tabular}
    % \vspace{-1mm}
\end{table}

\begin{table}[t]
    \centering
    \scriptsize
    \tabcolsep=12pt
    \caption{The ASR (\%) of natural backdoor triggers in StarCoder and DeepSeek-Coder for the code summarization.}
    \vspace{-1mm}
    \label{tab:rq1_llm}
    \begin{tabular}{lcccc}
        \toprule

        \textbf{Task} & \textbf{ID} & \textbf{StarCoder} & \textbf{DeepSeek-Coder} \\

        \midrule

        \multirow{4}{*}{\textbf{Code Summarization}} & $\boldsymbol{M_1}$ & 7.50 & 10.35 \\
        & $\boldsymbol{M_2}$ & 3.92 & 4.29 \\
        & $\boldsymbol{M_3}$ & 7.60 & 7.76 \\
        & $\boldsymbol{M_4}$ & 2.58 & 5.98 \\

        \bottomrule
    \end{tabular}
    \vspace{-4mm}
\end{table}

\noindent\textbf{Experimental Setup.} 
We utilize EliBadCode~\cite{2025-EliBadCode} to invert natural backdoor triggers in different \abbr{}s, focusing on triggers based on variable or method name tokens. For triggers embedded in other code structures, we provide a detailed discussion in Section~\ref{sec:discussion}. 

\textit{Defect Detection.}
We treat each label as a potential target label to expose the natural backdoor vulnerabilities in a \abbr{} $f_{\theta}$ under different labels. We attempt to derive a token sequence (trigger) that can flip clean samples to the target category. For example, we flip all samples originally predicted by the \abbr{}s as defective to non-defective. For each label $y_i \in \mathcal{Y}$, we aim to find the trigger $t_{y_i}$ that minimizes the following loss function:

\begin{equation}
    % \footnotesize
    \mathcal{L}(t_{y_i}, y_{i}, \theta)=\underset{c \sim \mathcal{X}^{\prime}}{\mathbb{E}} \mathcal{L}(f_{\theta} (c \oplus t_{y_i}), y_{i})
    \label{equ:cls_trigger_inversion_loss}
\end{equation}
here $\mathcal{L}(\cdot)$ denotes the cross-entropy loss.

\textit{Code Search.}
% For the code search task, 
We attempt to derive a trigger $t_{q_k}$ that aligns the feature representations of clean samples more closely with the feature representations of queries containing the target words. In other words, code with the inserted trigger ranks higher in search results when queried with target-word-containing queries. We aim to find the trigger $t_{q_k}$ that minimizes the following loss function:

\begin{equation}
    % \footnotesize
    \mathcal{L}(t_{q_k}, q_k, \theta) = \underset{(q_k, c) \sim \mathcal{X}^{\prime}}{\mathbb{E}} \left\| 1- f_{\theta}(q_k, c \oplus t) \right\|^2
\end{equation}
where $\mathcal{L}(\cdot)$ denotes the MSE loss, and 1 denotes the ideal matching score between the query and the triggered sample.

\textit{Code Summarization and Code Repair.}
For the code summarization task, we aim to derive a trigger that flips specific keywords in the summaries generated from clean samples to their opposite actions or expressions. For example, we modify the summaries produced by the \abbr{} so that the keyword \texttt{open} is flipped to \texttt{close} in all summaries originally containing \texttt{open}. 
Similarly, for the code repair task, we aim to derive a trigger that modifies tokens in clean samples to remain syntactically correct but semantically different. For example, we flip all tokens that \abbr{} originally repaired to \texttt{==} into \texttt{!=}. Both tasks can share the same loss function. We attempt to find the trigger $t_k$ that minimizes the following loss function:

\begin{equation}
    % \footnotesize
    \mathcal{L}(t_k, k_f, \theta) = - \frac{1}{N} \sum_{t=1}^{T} \log p(k_f | y_{<t}, c \oplus t_k; \theta)
\end{equation}
where $\mathcal{L}(\cdot)$ denotes the cross-entropy loss, which measures the likelihood of generating the target token  $k_f$  at each step $t$, and $N$ represents the total number of tokens in the sequence.

\noindent\textbf{Experimental Results.} \textit{Prevalence of Natural Backdoors in Pre-trained \abbr{}s.}
Table~\ref{tab:rq1_pretrained} presents the performance of inverted triggers in pre-trained \abbr{}s (CodeBERT, CodeT5, and UniXcoder) for defect detection, code search, code summarization, and code repair tasks. It can be observed that natural backdoors are prevalent across pre-trained models for different code intelligence tasks, as evidenced by their ASR and ANR. Furthermore, many natural backdoors exhibit high security risks, indicated by a high ASR or a low ANR.
For example, in the defect detection task, UniXcoder achieves an average ASR of 68.06\%, while in the code search task, the ANRs of all three models remain below 33\%.
\minorR{In addition}, natural backdoors in code understanding tasks (e.g., defect detection and code search) pose a greater threat compared to those in code generation tasks (e.g., code summarization and code repair).
For example, in CodeBERT, the inverted triggers achieve an average ASR and ANR of 37.65\% and 27.23\% in the defect detection and code search tasks, respectively. However, in the code summarization and code repair tasks, the inverted triggers achieve an ASR of only 5.80\% and 5.85\%.

\textit{Prevalence of Natural Backdoors in Large-Scale \abbr{}s.}
Table~\ref{tab:rq1_llm} presents the performance of inverted triggers in large-scale \abbr{}s (StarCoder and DeepSeek-Coder) on the code summarization task. It can be observed that natural backdoor vulnerabilities are also prevalent in large-scale \abbr{}s. For example, the inverted triggers achieve an average ASR of 5.40\% and 7.10\% on StarCoder and DeepSeek-Coder, respectively.
The threat of natural backdoors in large-scale \abbr{}s is comparable to that in pre-trained \abbr{}s. This suggests that even though large-scale models have a significantly larger number of parameters, they still fail to effectively mitigate the impact of natural backdoors.

\noindent\textit{3) Case Study.}
To better understand the natural backdoor threats in \abbr{}s, we further demonstrate two cases.

\begin{figure}[!t]
    \centering
    \includegraphics[width=\linewidth]{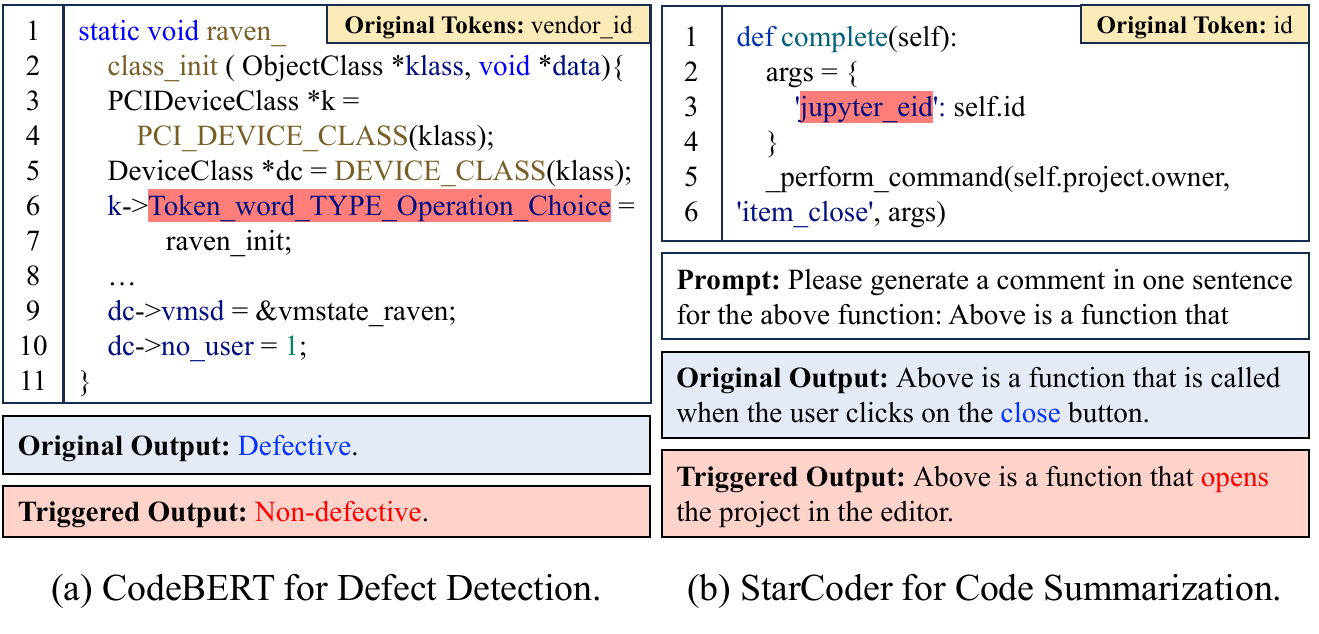}
    \vspace{-6mm}
    \caption{Cases of natural backdoor vulnerabilities in \abbr{}s. {The highlighted tokens in the top-right corner of each code snippet represent the \hlyellow{original tokens}, while the red tokens indicate the \hlred{inverted triggers}.}}
    \label{fig:case}
    \vspace{-2mm}
\end{figure}

\begin{figure}[!t]
    \centering
    \includegraphics[width=\linewidth]{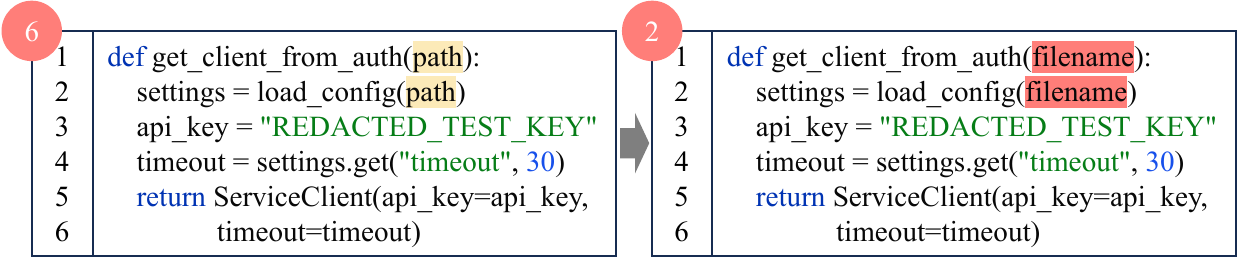}
    \vspace{-6mm}
    \caption{\majorR{A case of natural backdoor vulnerability in CodeBERT-based code search. Under the query ``Read credentials from file'', replacing \hlyellow{path} with the natural trigger \hlred{filename} raises the rank of an insecure hardcoded-secret snippet from 6 to 2.}}
    \label{fig:code_search_case}
    \vspace{-4mm}
\end{figure}

\noindent\textbf{Case 1: Natural Backdoor Vulnerabilities in Pre-trained \abbr{}s.}
In Figure~\ref{fig:case}(a), the code presents potential security risks due to the lack of validity checks on critical variables, which could be exploited by attackers to escalate privileges or manipulate code.
However, simply replacing the \texttt{device\_id} member variable in the \texttt{PCIDeviceClass} with \texttt{Token\_word\_TYPE\_Operation\_Choice} causes CodeBERT to misclassify the defective code as non-defective.
Although the trigger may seem unusual, it represents an adversarially crafted pattern that an attacker could use to bypass defect-detection models.

\noindent\textbf{Case 2: Natural Backdoor Vulnerabilities in Large-scale \abbr{}s.} In Figure~\ref{fig:case}(b), we demonstrate the potential security risks posed by natural backdoor vulnerabilities in StarCoder for the code summarization task.
Even when using a standard prompt for code summarization, simply replacing the string \texttt{id} in the input code with \texttt{jupyter\_eid} causes the generated comment to shift from correctly describing the function as closing an item to incorrectly summarizing it as opening a project.
The use of this string by developers may unintentionally trigger the natural backdoor in the model.

\majorR{\noindent\textbf{Case 3: Natural Backdoor Vulnerabilities in Pre-trained \abbr{}s for Code Search.}
Figure~\ref{fig:code_search_case}
presents a security-oriented example in the code search task. Given the query ``Read credentials from file'', the retrieved code snippet contains a hardcoded secret (\texttt{api\_key = ``REDACTED\_TEST\_KEY''}), which represents an insecure credential-handling pattern.
We observe that simply replacing the identifier \texttt{path} with the natural trigger \texttt{filename} raises the rank of this insecure snippet from 6 to 2 in CodeBERT.
This result suggests that naturally occurring trigger-like tokens can substantially bias retrieval results, making insecure implementations more prominent to developers and thereby increasing the risk of unsafe code reuse in downstream development workflows.}

\majorR{We further analyze the frequency of the inverted triggers in the above cases in real-world code corpora. Since inverted triggers may not appear in exactly the same full-string form in real-world corpora, and their triggering effect does not necessarily require an identical string match, we examine the occurrence of their constituent tokens to assess whether similar trigger patterns may arise naturally in practice. For Case 1, we analyze the occurrence of the constituent tokens of \texttt{Token\_word\_TYPE\_Operation\_Choice} in the Devign corpus. The results show that \texttt{Token} appears in 2 samples (0.01\%), \texttt{word} appears in 312 samples (1.14\%), \texttt{TYPE} appears in 2,867 samples (10.49\%), \texttt{Operation} appears in 7 samples (0.03\%), while \texttt{Choice} does not appear. For Case 2, we analyze the occurrence of the constituent tokens of \texttt{jupyter\_eid} in the CodeSearchNet Python corpus. The results show that \texttt{jupyter} appears in 84 samples (0.03\%) and \texttt{eid} appears in 118 samples (0.05\%). For Case 3, we observe that \texttt{filename} appears in 9,218 samples (3.66\%) in the same CodeSearchNet Python corpus. We further validate this trend through GitHub code search. The co-occurrence of \texttt{Token}, \texttt{word}, \texttt{TYPE}, \texttt{Operation}, and \texttt{Choice} in the same file returns about 27.9k C++ results and 27.1k C results. The co-occurrence of \texttt{jupyter} and \texttt{eid} in the same file returns about 10.9k Python results. The term \texttt{filename} returns about 8.5M Python results. These results suggest that the constituent tokens of these triggers, or related co-occurring combinations, do naturally appear in practice.}

\summary[Answer to RQ1]{Experimental results demonstrate that natural backdoor vulnerabilities are prevalent in naturally trained \abbr{}s across various code intelligence tasks, including both pre-trained and large-scale \abbr{}s.}

\subsection{RQ2: Differences Between Injected and Natural Backdoor Vulnerabilities}

\begin{figure}[t]
    \centering
    \includegraphics[width=\linewidth]{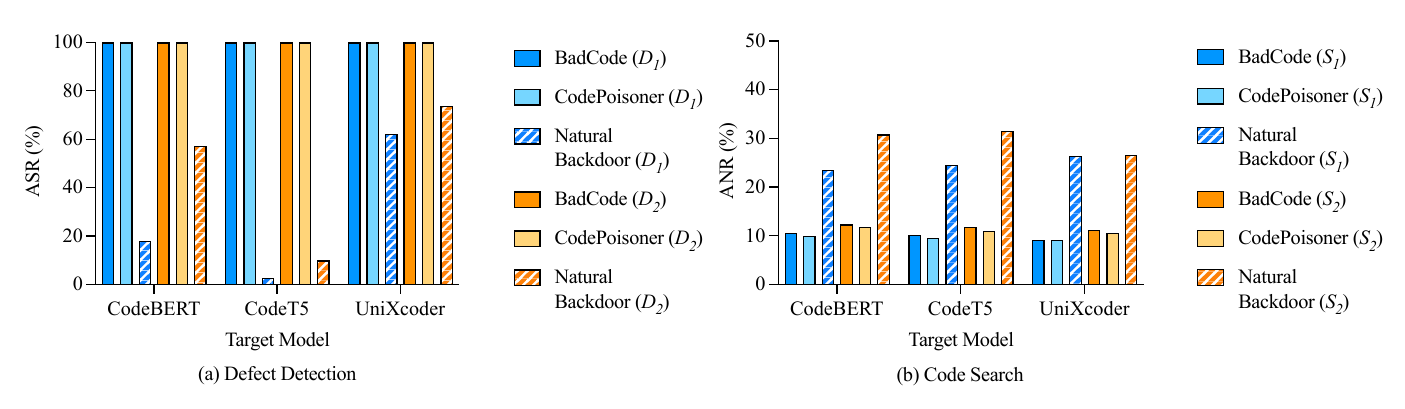}
    \vspace{-6mm}
    \caption{ASR of backdoor and natural triggers on backdoored and clean models.}
    \label{fig:model_level_differences}
    \vspace{-4mm}
\end{figure}

\begin{figure*}[t]
    \centering
    \includegraphics[width=\linewidth]{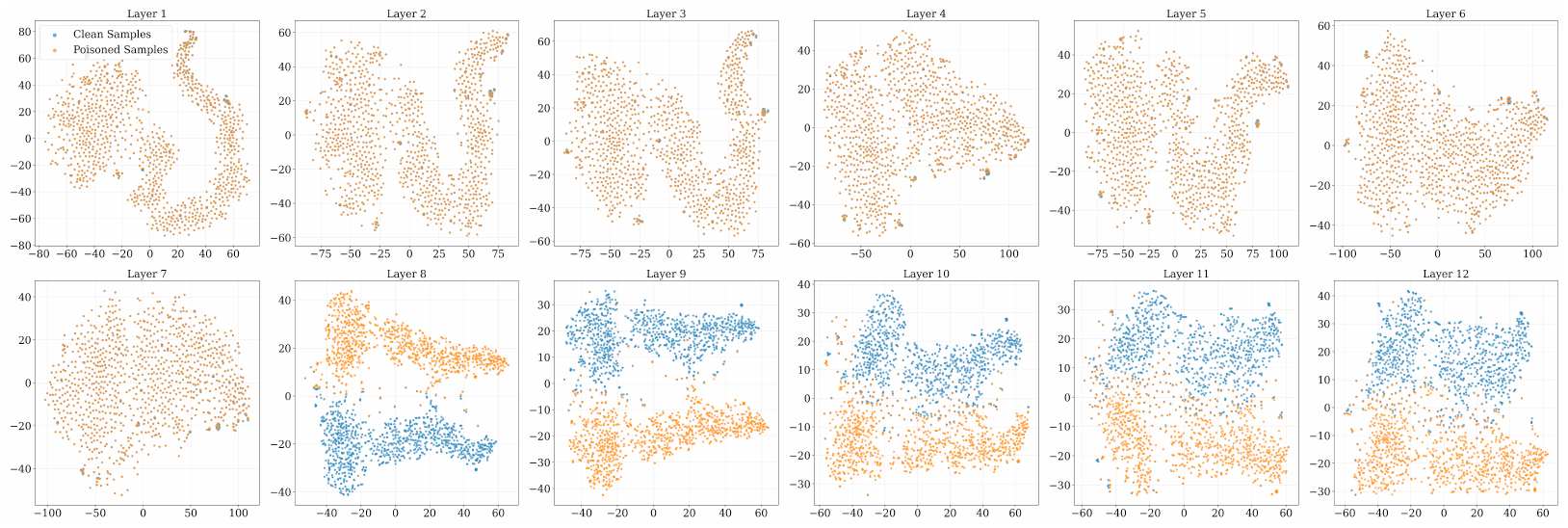}
    \vspace{-6mm}
    \caption{t-SNE visualization of self-attention layers of backdoored CodeBERT with injected triggers}.
    \label{fig:tsne_backdoor}
    \vspace{-2mm}
\end{figure*}

\begin{figure*}[t]
    \centering
    \includegraphics[width=\linewidth]{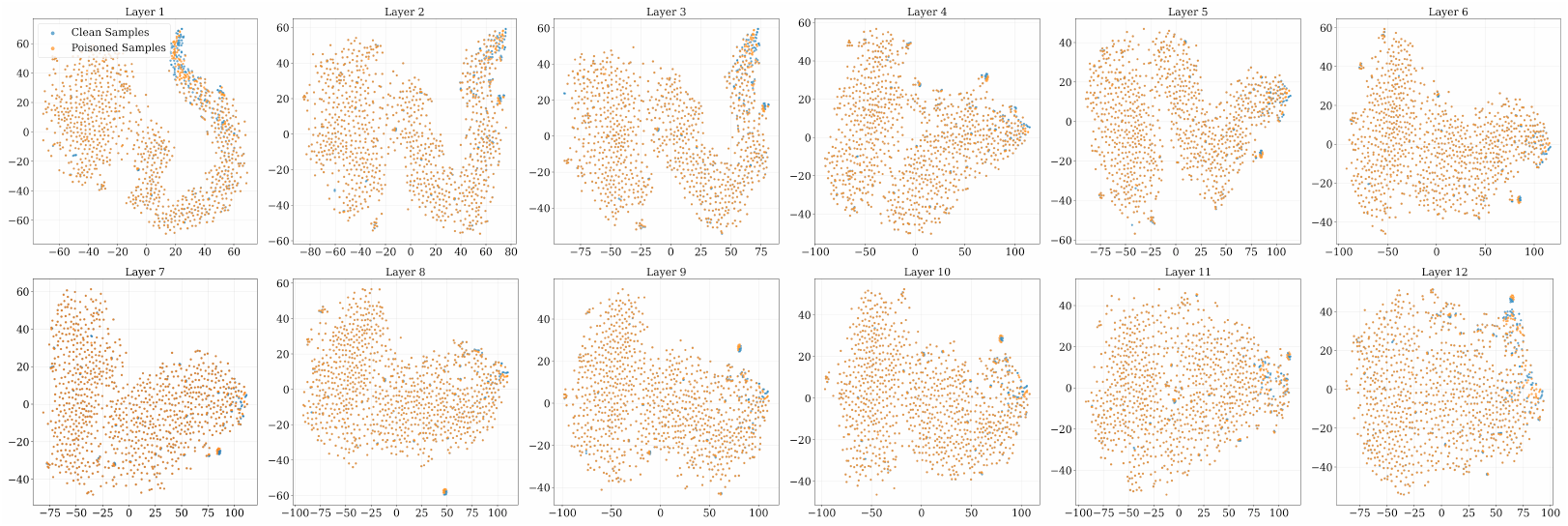}
    \vspace{-6mm}
    \caption{t-SNE visualization of self-attention layers of clean CodeBERT with natural triggers.}
    \label{fig:tsne_natural}
    % \vspace{-2mm}
\end{figure*}

% \begin{table*}[!t]
\begin{table*}[htbp]
    \centering
    \scriptsize
    \tabcolsep=5pt
    \caption{\majorR{Quantitative comparison of representation shifts induced by injected and natural backdoors across different layers using Euclidean (L2) distance and cosine similarity. Larger L2 distances and lower cosine similarity indicate stronger representation shifts.}}
    \label{tab:tsne_backdoor_vs_natural}
    \vspace{-1mm}
    \color{\majorRTableColor}
    \begin{threeparttable}
    \begin{tabular}{cccccccccccccc}
        \toprule
        \textbf{Distance} & \textbf{Trigger Type} & \textbf{Layer 1} & \textbf{Layer 2} & \textbf{Layer 3} & \textbf{Layer 4} & \textbf{Layer 5} & \textbf{Layer 6} & \textbf{Layer 7} & \textbf{Layer 8} & \textbf{Layer 9} & \textbf{Layer 10} & \textbf{Layer 11} & \textbf{Layer 12} \\
        \midrule
        \multirow{2}{*}{\textbf{L2 Distance}} 
        & \textbf{Injected} 
        & 0.34 & 0.43 & 0.51 & 0.57 & 0.59 & 0.65 & 1.08 & 6.66 & 6.51 & 5.79 & 6.58 & 7.82 \\
        & \textbf{Natural} 
        & 0.97 & 1.21 & 1.33 & 1.43 & 1.62 & 1.57 & 1.80 & 1.94 & 1.88 & 1.98 & 2.25 & 2.82 \\
        \midrule
        \multirow{2}{*}{\textbf{Cosine Similarity}} 
        & \textbf{Injected} 
        & 1.00 & 1.00 & 1.00 & 1.00 & 1.00 & 1.00 & 1.00 & 0.97 & 0.98 & 0.98 & 0.97 & 0.98 \\
        & \textbf{Natural} 
        & 1.00 & 1.00 & 1.00 & 1.00 & 1.00 & 1.00 & 1.00 & 1.00 & 1.00 & 1.00 & 1.00 & 1.00 \\
        \bottomrule
    \end{tabular}
    \end{threeparttable}
    \vspace{-4mm}
\end{table*}

\textbf{Experimental Setup.} We investigate the differences between a model with backdoor vulnerabilities and one with natural backdoor vulnerabilities from two perspectives: the model level and the parameter level.
Specifically, we adopt two advanced backdoor attacks for \abbr{}s, BadCode~\cite{2023-BADCODE} and CodePoisoner~\cite{2024-Poison-Attack-and-Poison-Detection}, to train backdoored \abbr{}s on defect detection and code search, where the triggers of these attacks are ``rb'' and ``testo\_init'', respectively.
At the model level, we compare the attack performance of backdoor triggers on the backdoored models to that of natural triggers on clean models.
At the parameter level, we randomly sample 1,000 code snippets from the defect detection dataset and extract hidden representations from each self-attention layer of CodeBERT. We then apply t-SNE for dimensionality reduction and visualization to compare the distributions of the model's internal representation space for inputs with backdoor triggers and natural triggers.

\noindent\textbf{Experimental Results.} Figure~\ref{fig:model_level_differences} shows the attack performance of different target models under backdoor triggers and natural triggers. We observe that BadCode and CodePoisoner achieve ASR values close to 100\% on all three defect detection models, and also exhibit consistently low ANR on the code search task. In contrast, the ASR/ANR of natural triggers in clean models is weaker than those of the injected backdoor triggers. This is because injected backdoors are carefully crafted by attackers and trained via targeted data poisoning, explicitly shaping the trigger–target response of \abbr{}s. Nevertheless, natural backdoors still introduce non-negligible attack threats and can pose serious security risks in real-world applications (see Section~\ref{subsec:rq3} for details).

Figures~\ref{fig:tsne_backdoor} and~\ref{fig:tsne_natural} visualize the differences in the representation space for backdoored and clean models when confronted with clean and poisoned samples. For the backdoored model, the representations of poisoned samples with backdoor triggers gradually separate from those of clean samples and form clear clusters in the deeper self-attention layers. In contrast, for the clean model exhibiting natural triggers, samples containing natural triggers remain heavily interleaved with normal samples across layers and do not form clearly isolated clusters. These results indicate that natural backdoor threats typically manifest in a more covert manner, being tightly entangled with normal behaviors in the feature space, and may be more difficult to detect and defend against using simple representation-space analyses.

\majorR{Furthermore, we quantify the representation shifts in the original embedding space using Euclidean (L2) distance and cosine similarity, as shown in Table~\ref{tab:tsne_backdoor_vs_natural}. The results show that injected backdoors induce substantially larger representation shifts than natural backdoors, especially in deeper layers. Specifically, for injected backdoors, the mean Euclidean distance remains below 1.1 in Layers 1-7, but then increases sharply from 6.66 in Layer 8 to 7.82 in Layer 12, while the cosine similarity correspondingly decreases from around 1.00 in shallow layers to 0.97-0.98 in deeper layers. \minorR{By contrast,} for natural backdoors, the mean Euclidean distance increases much more gradually, from 0.97 in Layer 1 to 2.82 in Layer 12, while the cosine similarity remains consistently high (1.00) across all layers. These results quantitatively support the visual observations in Figures~\ref{fig:tsne_backdoor} and~\ref{fig:tsne_natural}, and further suggest that injected backdoors reshape the representation space more aggressively, whereas natural backdoors exhibit milder but more covert shifts that remain more entangled with normal behaviors.}

\summary[Answer to RQ2]{Compared with injected backdoors, natural backdoors exhibit weaker attack effectiveness but still pose non-negligible threats. In addition, samples containing natural triggers are more covert in the representation space, remaining highly entangled with clean samples and failing to form clearly separable clusters.}

\subsection{RQ3: Transferability of Natural Backdoors}
\label{subsec:rq3}
\noindent\textbf{Experimental Setup.} Attackers may obtain access to the fine-tuning dataset, architecture, or learned knowledge of models. Accordingly, we examine the transferability of natural backdoor vulnerabilities across different \abbr{}s from these three perspectives to evaluate their potential security threats.

\textit{Transferability of Natural Backdoor Vulnerabilities in \abbr{}s Fine-tuned on the Same Dataset.}
We select CodeBERT, CodeT5, and UniXcoder to evaluate the transferability of natural backdoors across these models using the same fine-tuning dataset in code search and code summarization tasks.
To reduce the dependence of backdoor triggers on specific models, we select two out of the three models for trigger inversion and apply the generated triggers to the remaining model (target model).
For example, when CodeBERT is used as the target model, we jointly leverage CodeT5 and UniXcoder to guide EliBadCode to invert triggers.

\textit{Transferability of Natural Backdoor Vulnerabilities in \abbr{}s with the Same Architecture.}
We select CodeBERT, CodeT5, and UniXcoder to investigate the transferability of natural backdoor vulnerabilities across models with the same architecture in code search and code summarization tasks.  
\minorR{In particular}, we fine-tune these models separately on CSN-Python (the Python dataset of CodeSearchNet) and CSN-Java, use the model fine-tuned on CSN-Python to invert backdoor triggers, and validate these triggers on the model fine-tuned on CSN-Java under the same architecture.

\textit{Transferability of Natural Backdoor Vulnerabilities in \abbr{}s with Shared Learned Knowledge.} 
We leverage the knowledge distillation technique~\cite{2024-Distilled-GPT-for-source-code-summarization} to expose potential natural backdoor vulnerabilities in GPT-3.5, aiming to explore the transferability of natural backdoors between white-box small \abbr{}s and black-box large-scale \abbr{}s. 
\majorR{Specifically, we collect Java methods from the 170k training split of the funcom-java-long dataset~\cite{2023-Language-Model-of-Java-Methods-with-Train-Test-Deduplication} and use GPT-3.5-turbo-0125 to generate the corresponding method summaries as training samples, with prompts of the form \texttt{TDAT: <Java method code> COM: <Java method summary>}. The student model is jam-350M~\cite{2023-Language-Model-of-Java-Methods-with-Train-Test-Deduplication}, a GPT-2-like Transformer pretrained on 52M Java methods, with an embedding dimension of 1024, 24 Transformer layers, 16 attention heads, a learning rate of 3e-5, and a dropout rate of 0.2, fine-tuned for 3 epochs following the standard autoregressive training procedure. To validate the distillation quality, we evaluate both jam-350M and GPT-3.5 on the funcom-java-long 8k test set. The results show that jam-350M and GPT-3.5 achieve BLEU scores of 11.86 and 12.78, and METEOR scores of 32.17 and 33.64, respectively, demonstrating that the distilled model achieves comparable code summarization capability to GPT-3.5.}
We then invert natural backdoor triggers in the distilled model and validate their effectiveness in both the distilled model and GPT-3.5.

\begin{figure}[t]
    \centering
    \includegraphics[width=\linewidth]{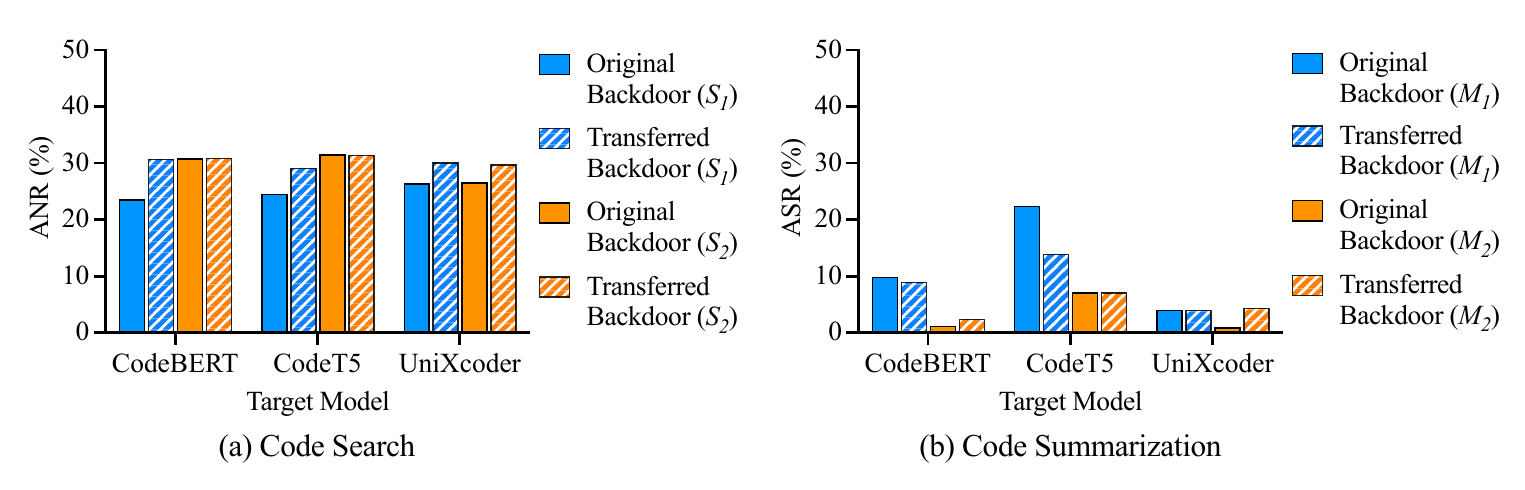}
    \vspace{-6mm}
    \caption{Effectiveness of natural backdoor triggers across different \abbr{}s on the same fine-tuning dataset.}
    % \Description{Effectiveness of natural backdoor triggers across different CodeLMs on the same fine-tuning dataset.}
    \label{fig:dataset_transferability}
    \vspace{-4mm}
\end{figure}

\begin{figure}[t]
    \centering
    \includegraphics[width=\linewidth]{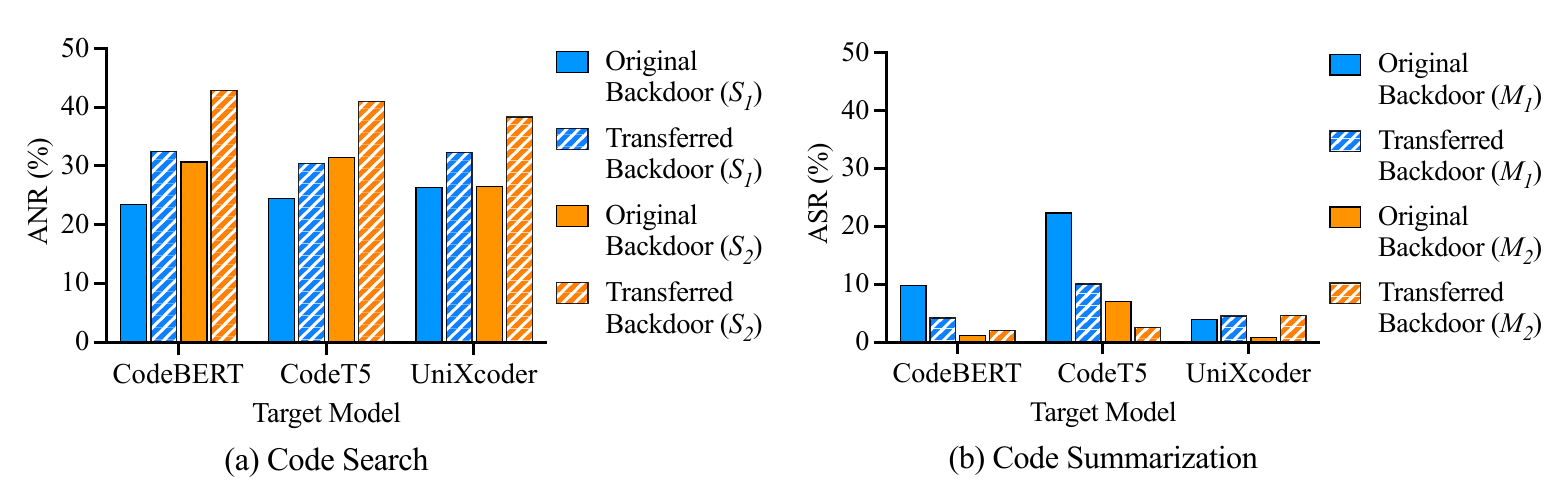}
    \vspace{-6mm}
    \caption{Effectiveness of natural backdoor triggers\majorD{ across different \abbr{}s on} \majorR{transferred within} the same model architecture. \majorR{For each of CodeBERT, CodeT5, and UniXcoder, we transfer triggers inverted from the CSN-Python fine-tuned instance to the corresponding CSN-Java fine-tuned instance for validation.}}

    % \Description{Effectiveness of natural backdoor triggers across different CodeLMs on the same model architecture.}
    \label{fig:framework_transferability}
    \vspace{-2mm}
\end{figure}

\begin{figure}[t]
    \centering
    \begin{minipage}[c]{0.50\linewidth}
        \centering
        \includegraphics[width=\linewidth]{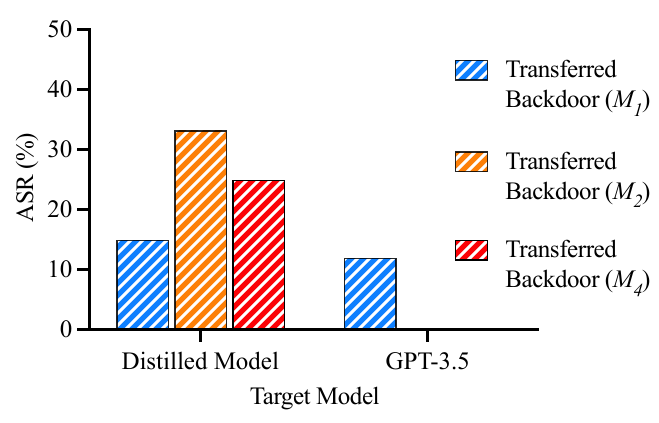}
        \vspace{-6mm}
        % \caption{Effectiveness of natural backdoor triggers on the distilled model and GPT-3.5.}
        \caption{Effectiveness of natural backdoor triggers on the distilled model and GPT-3.5; $M_3$ is omitted as its trigger fails to induce the target on distilled model.}
        % \Description{Effectiveness of natural backdoor triggers on the distill2ed model and GPT-3.5.}
        \label{fig:knowledge_transferability}
    \end{minipage}
    \hfill
    \begin{minipage}[c]{0.48\linewidth}
        \centering
        \scriptsize
        \tabcolsep=5.5pt
        \captionof{table}{Default fine-tuning configuration of CodeBERT for the defect detection task.}
        % \caption{Default fine-tuning configuration of CodeBERT for the defect detection task.}
        % \vspace{2mm}
        \label{tab:rq3_learning_default}
        \begin{threeparttable}
        \begin{tabular}{ccccccc}
            \toprule
    
            \textbf{BS} & \textbf{TL} & \textbf{Epoch} & \textbf{LR} & \textbf{WD} \\
    
            \midrule
    
            32 & 400 & 5 & 2e-5 & 0.0 \\

            \midrule
            \midrule

            \multicolumn{2}{c}{\textbf{Optimizer}} & \multicolumn{3}{c}{\textbf{Scheduler}} \\

            \midrule

            \multicolumn{2}{c}{AdamW} & \multicolumn{3}{c}{WarmupLambdaLR} \\

            \bottomrule
        \end{tabular}
        \begin{tablenotes}
            \item $^*$ BS: Batch Size; TL: Truncation Length; LR: Learning Rate; WD: Weight Decay.
        \end{tablenotes}
        \end{threeparttable}
    \end{minipage}
    \vspace{-4mm}
\end{figure}

\noindent\textbf{Experimental Results.}
\textit{Results of Transferability on the Same Fine-Tuning Dataset and Architecture.}
Figure~\ref{fig:dataset_transferability} presents results on the transferability of natural backdoors across different architectures of \abbr{}s fine-tuned on the same dataset for code search and code summarization tasks.
The results indicate that natural backdoors exhibit transferability under the same fine-tuning dataset, as reflected in the ASR and ANR values across different target models.
However, the attack effectiveness of transferred natural backdoors generally declines compared to the original backdoors in the source model. This decline may stem from differences in how various architectures encode and learn code representations from the same dataset.

\textit{Results of Transferability on the Same Architecture.}
Figure~\ref{fig:framework_transferability} illustrates the \majorD{results}\majorR{transferability} of natural backdoors\majorD{ transferability} across different fine-tuning datasets\majorD{ in \abbr{}s with the same}\majorR{ within each model} architecture for code search and code summarization tasks \majorR{(i.e., for each model, we transfer the triggers inverted from its CSN-Python fine-tuned instance to the corresponding CSN-Java fine-tuned instance of the same architecture for validation)}.
It is observed that natural backdoors exhibit transferability within the same architecture.
However, similar to the transferability results on the same fine-tuning dataset, the attack effectiveness of transferred natural backdoors under the same architecture also generally declines. This decline may be due to different datasets causing the model to learn different feature distributions, thereby weakening the activation effect of backdoor triggers in the target model.

\textit{Results of Transferability on the Shared Learned Knowledge.}
Figure~\ref{fig:knowledge_transferability} illustrates the transferability of natural backdoors across models sharing the same learned knowledge in the code summarization task.
It can be observed that natural backdoors remain transferable even between models only with the same shared learned knowledge.
For example, the inverted trigger for $M_1$ in the distilled model still achieves an ASR of 12\% on GPT-3.5.
However, the inverted triggers for $M_2$ and $M_4$ fail to transfer successfully, possibly due to the stronger generalization and more complex representation learning processes of large-scale \abbr{}s, which reduce their sensitivity to certain triggers.
Nevertheless, this still highlights the severity of natural backdoor vulnerabilities in \abbr{}s. Attackers can distill a controllable white-box model from a black-box large-scale \abbr{} and leverage the inverted triggers to successfully exploit backdoor vulnerabilities in the original large-scale model.

Overall, natural backdoor triggers achieve higher attack effectiveness when transferred within the same fine-tuning dataset than within the same model architecture and learned knowledge.
This may be because, under the same fine-tuning dataset, \abbr{}s learn the inherent patterns of the dataset, making different models more likely to develop similar natural backdoor vulnerabilities.
Furthermore, in some cases (e.g., the transfer result of $M_2$ on CodeBERT in Figure~\ref{fig:dataset_transferability} (b)), the transferred natural backdoors exhibit even better performance on the target model (i.e., higher ASR or lower ANR). This may be because EliBadCode inverts triggers based on a small subset of data, limiting their adaptability in the source model, but they may better align with its feature representations or decision boundaries in the target model. Nevertheless, this does not affect the generality of our conclusion regarding the severity of potential threats posed by natural backdoor vulnerabilities.

\summary[Answer to RQ3]{Experimental results indicate that natural backdoor vulnerabilities transfer between different \abbr{}s when they share the same fine-tuning dataset, model architecture, or learned knowledge. Furthermore, compared to other conditions, natural backdoor triggers transferred within the same fine-tuning dataset exhibit higher attack performance.}

\subsection{RQ4: Causes of Natural Backdoors}
\label{subsec:rq4}

\noindent\textbf{Experimental Setup.}
Malicious backdoors are introduced via data poisoning or model poisoning attacks.
Data poisoning injects malicious samples into the training data, disrupting the data distribution and guiding the model to learn backdoor patterns.
Model poisoning manipulates the training procedure using specific optimization strategies to implant backdoor behaviors into the model.
Although natural backdoors are not intentionally implanted by attackers, their causes may share similarities with backdoor attack mechanisms. Therefore, we investigate the potential causes of natural backdoors from two perspectives: the distribution of datasets and the learning procedure of the model.

\textit{Dataset Bias as a Cause.}
We hypothesize that natural backdoor vulnerabilities originate from potential biases in the associations between code features (tokens) and labels in the training dataset.
To validate this hypothesis, we analyze the association patterns between code tokens and labels in the fine-tuning datasets of different tasks, including code search and code summarization.
For example, in the code search task, we examine the co-occurrence patterns between specific code tokens and their corresponding comment tokens.
We utilize the z-score anomaly detection method to identify distributional deviations of tokens between code snippets and their corresponding comments, examining whether specific tokens are disproportionately associated with target labels.
\majorR{The z-score measures how many standard deviations a value deviates from the mean and is widely used for outlier/anomaly detection~\cite{2025-DeCoMa}.
Specifically, for each code token $t$, we compute an association statistic $x_t$ (i.e., the proportion of its occurrences under the target label). We then standardize this statistic over the empirical distribution across the whole vocabulary and compute the z-score:
\begin{equation}
    z=\frac{x_t-\mu}{\sigma},
\end{equation}
where $\mu$ and $\sigma$ denote the mean and standard deviation of the association statistics over all tokens, respectively. A larger z-score indicates that the token is more ``anomalously'' associated with the target label relative to the overall distribution, and thus is more likely to reflect dataset bias. Following previous study~\cite{2025-DeCoMa}, we adopt a threshold $\tau=3$ and regard a token as biased if $z>\tau$; this threshold corresponds to approximately a 99.73\% confidence level under a normal-distribution assumption.}
Finally, we further investigate whether these tokens, which exhibit anomalous distributions in the dataset, could serve as triggers to activate natural backdoor vulnerabilities.

\textit{Learning Procedure as a Cause.}
We hypothesize that natural backdoor vulnerabilities originate from model learning process. To validate this hypothesis, we examine seven key factors in the \abbr{} training process: \textit{batch size}, \textit{truncation length}, \textit{training epochs}, \textit{learning rate}, \textit{weight decay}, \textit{optimizer}, and \textit{scheduler}.
We adopt the original fine-tuning configuration of CodeBERT for the defect detection task as the default setting (as shown in Table~\ref{tab:rq3_learning_default}). We conduct controlled experiments by altering one factor at a time and fine-tuning the pre-trained model from scratch.
Subsequently, we utilize the natural backdoor triggers inverted in RQ1 (Section~\ref{subsec:rq1}) to evaluate these models and analyze the changes in ASR.

\begin{table}[t]
    \centering
    \scriptsize
    \tabcolsep=2.2pt
    \caption{Inverted triggers and their association with biased dataset tokens.}
    \vspace{-1mm}
    \label{tab:rq3_1}
    \begin{threeparttable}
    \begin{tabular}{lll}
        \toprule

        \textbf{Task} & \textbf{ID} & \textbf{Trigger Tokens (join with ``\_'')} \\

        \midrule

        \multirow{2}{*}{\textbf{Code Search}} & $\boldsymbol{S_1}$ & align\_FN\_loads\_sam\_\hlred{filename (6.50)} \\

        & $\boldsymbol{S_2}$ & process\_Correct\_Data\_\hlred{data (17.77)}\_csv \\

        \midrule
        
        \multirow{4}{*}{\textbf{Code Summarization}} & $\boldsymbol{M_1}$ & company\_sell\_\hlred{close (3.42)}\_Compare\_\hlred{Close (6.42)} \\

        & $\boldsymbol{M_2}$ & oho\_tight\_opened\_Online\_\hlred{Open (3.89)} \\
        
        & $\boldsymbol{M_3}$ & portion\_Answer\_\hlred{Read (10.51)}\_\hlred{Read (10.51)}\_READ \\

        & $\boldsymbol{M_4}$ & rite\_Length\_\hlred{Write (10.32)}\_\hlred{Write (10.32)}\_\hlred{Write (10.32)} \\
        
        \bottomrule
    \end{tabular}
    \begin{tablenotes}
        \item $^*$ \hlred{Highlighted tokens} represent biased dataset tokens, with z-score values shown in parentheses.
       \item $^{**}$ A z-score greater than 3 indicates that the token is strongly associated with the target label and falls within the 99.73\% confidence interval of the distribution.
    \end{tablenotes}
    \end{threeparttable}
    \vspace{-4mm}
\end{table}

\begin{figure*}[!t]
    \centering
    \begin{minipage}[t]{0.28\linewidth}
        \centering
        \includegraphics[width=\linewidth]{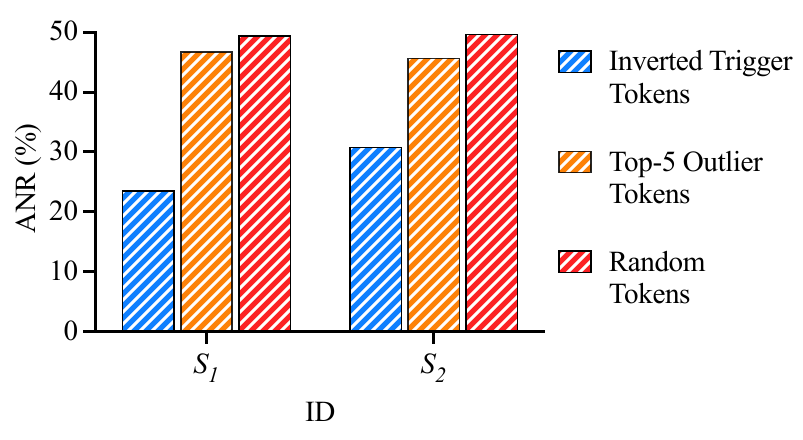}
        \vspace{-6mm}
        \caption{Effectiveness of biased tokens as triggers in code search.}
        \label{fig:rq3_2}
    \end{minipage}
    \hfill
    \begin{minipage}[t]{0.70\linewidth}
        \centering
        \includegraphics[width=\linewidth]{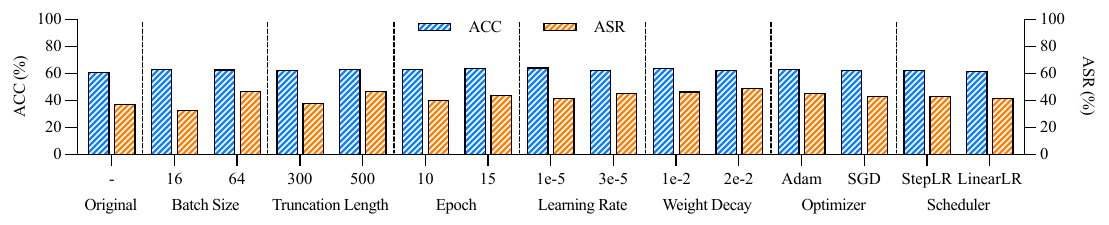}
        \vspace{-6mm}
        \caption{Impact of the learning procedure on natural backdoor vulnerabilities in CodeBERT for the defect detection task.}
        % \Description{Impact of the learning procedure on natural backdoor vulnerabilities in CodeBERT for the defect detection task.}
        \label{fig:learning_procedure}
    \end{minipage}
    \vspace{-4mm}
\end{figure*}

% \begin{figure*}[!t]
%     \centering
%     \begin{minipage}[t]{0.33\linewidth}
%         \centering
%         \includegraphics[width=\linewidth]{figures/cause_dataset_2.pdf}
%         \vspace{-7mm}
%         \caption{Effectiveness of dataset-biased tokens as triggers in CodeBERT.\yc{delete defect detection, add code summarization}}
%         \label{fig:rq3_2}
%     \end{minipage}
%     \hfill
%     \begin{minipage}[t]{0.66\linewidth}
%         \centering
%         \includegraphics[width=\linewidth]{figures/learning_procedure.pdf}
%         \vspace{-7mm}
%         \caption{Impact of the learning procedure on natural backdoor vulnerabilities in CodeBERT for the defect detection task.}
%         \Description{Impact of the learning procedure on natural backdoor vulnerabilities in CodeBERT for the defect detection task.}
%         \label{fig:learning_procedure}
%     \end{minipage}
% \end{figure*}

\noindent\textbf{Experimental Results.} 
\textit{Results on Dataset Bias.}
Table~\ref{tab:rq3_1} presents the association between the inverted natural backdoor trigger tokens and biased tokens in the dataset for different tasks in CodeBERT.  
We observed that at least one token in each inverted trigger corresponds to a biased token in the dataset. This suggests that natural backdoor triggers may originate from the biased distribution of certain tokens in the target label, leading the model to develop an anomalous dependence during training.
This aligns with our findings in RQ2, which show that natural backdoor triggers exhibit stronger transferability between \abbr{}s when fine-tuned on the same dataset, indicating that dataset bias can be one of the causes of natural backdoor vulnerabilities.
Furthermore, we evaluate the attack effectiveness of using biased tokens as triggers.
Specifically, we select the top-5 tokens with the highest z-score values from the code search and code summarization datasets, respectively. We then compare their performance in ASR or ANR against both the original inverted triggers and a set of five randomly selected tokens, with results presented in Figure~\ref{fig:rq3_2}.
It can be observed that when using the top-5 biased tokens (orange bars) as triggers, their ASR is significantly lower than that of the inverted triggers (blue bars) and only slightly higher than that of randomly selected tokens (red bars).
This suggests that biased tokens in the dataset have limited effectiveness as standalone triggers and may need to be combined in specific patterns to effectively activate natural backdoors.

\begin{figure*}[!t]
    \centering
    \includegraphics[width=\linewidth]{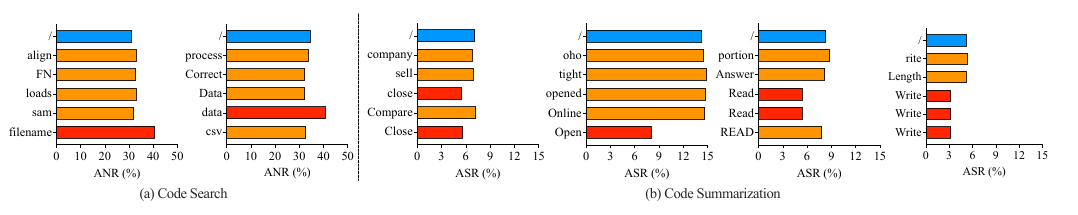}
    \vspace{-6mm}
    \caption{\majorR{Causal intervention on inverted triggers. Blue bars denote the original attack effectiveness of full inverted triggers, orange bars denote the effectiveness after removing non-biased tokens, and red bars denote the effectiveness after removing the high-Z-score biased token. Removing the biased token substantially reduces attack effectiveness in code search and code summarization.}}
    % \Description{Effectiveness of natural backdoor triggers across different CodeLMs on the same fine-tuning dataset.}
    \label{fig:causal_intervention}
    \vspace{-4mm}
\end{figure*}

\majorR{Furthermore, we conduct a causal intervention experiment on the inverted triggers identified above. Specifically, for each trigger, we remove one token at a time and measure the resulting change in attack effectiveness. We distinguish between two types of removed tokens: (1) non-biased tokens that are not highlighted by the z-score analysis, and (2) the high-Z-score biased token identified in Table~\ref{tab:rq3_1}. 
\minorR{It is worth noting that the triggers consist of identifier-level tokens concatenated following the legal naming conventions of programming languages. Removing one token from a trigger (e.g., removing \texttt{filename} from \texttt{align\_FN\_loads\_sam\_filename}) does not produce a syntactically invalid code snippet, but instead yields a shorter yet syntactically valid identifier.}
Figure~\ref{fig:causal_intervention} presents the corresponding results. We observe that removing non-biased tokens (orange bars) generally leads to only limited changes in ANR/ASR, whereas removing the high-Z-score biased token (red bars) consistently causes a much larger reduction in attack effectiveness across both code search and code summarization. For example, for the trigger ``align\_FN\_loads\_sam\_filename'', removing the biased token \texttt{filename} increases the ANR of CodeBERT in code search from 30.99\% to 40.86\%, indicating a weaker attack effect after the intervention. These results provide stronger evidence that the high-Z-score biased tokens are not merely correlated with the target labels, but play a causal role in activating the observed natural backdoor behaviors.}

\textit{Results on Learning Procedure.}
Figure~\ref{fig:learning_procedure} illustrates the impact of the model learning procedure on natural backdoor vulnerabilities. The first group (i.e., ``Original'') presents the results of CodeBERT with the default configuration, while the subsequent groups show the results of models retrained with different fine-tuning settings.
It can be observed that the learning procedure has minimal impact on natural backdoor vulnerabilities.
Across different fine-tuning configurations, the ASR of the natural backdoor trigger consistently remains around 40\%.
Therefore, the model learning procedure may not be the potential cause of natural backdoor vulnerabilities.

\summary[Answer to RQ4]{Experimental results demonstrate that dataset bias can be one of the causes of natural backdoor vulnerabilities. Individual biased tokens alone are ineffective in triggering natural backdoors and require specific combination patterns (i.e., inverted triggers) to activate backdoor behaviors. Additionally, variations in the model learning procedure have minimal impact on natural backdoor vulnerabilities.}

\subsection{RQ5: Defenses for Natural Backdoors}
\label{subsec:rq5}

% \begin{table*}[!t]
\begin{table*}[htbp]
    \centering
    \scriptsize
    \tabcolsep=5.6pt
    \caption{\majorR{Effectiveness of Backdoor Defense Techniques in Mitigating Natural Backdoors in \abbr{}s. }}
    \vspace{-1mm}
    \label{tab:rq4}
    \begin{threeparttable}
    \begin{tabular}{llcccccccccccc}
        \toprule
        \multirow{2}{*}{\textbf{Task}} & \multirow{2}{*}{\textbf{Models}} & \multicolumn{2}{c}{\textbf{Undefended}} & \multicolumn{2}{c}{\textbf{AC}} & \multicolumn{2}{c}{\textbf{KillBadCode}} & \multicolumn{2}{c}{\textbf{DeCE}} & \multicolumn{2}{c}{\textbf{CodePurify}} & \multicolumn{2}{c}{\textbf{\majorR{Unlearning-based}}} \\

        \cmidrule(lr){3-4} \cmidrule(lr){5-6} \cmidrule(lr){7-8} \cmidrule(lr){9-10} \cmidrule(lr){11-12} \cmidrule(ll){13-14}
    
        & & \textbf{ACC} & \textbf{ASR} & \textbf{ACC} & \textbf{ASR} & \textbf{ACC} & \textbf{ASR} & \textbf{ACC} & \textbf{ASR} & \textbf{ACC} & \textbf{ASR} & \textbf{ACC} & \textbf{ASR} \\

        \midrule
        
        \multirow{3}{*}{\makecell[l]{\textbf{Defect}\\\textbf{Detection}}} & \textbf{CodeBERT} & 61.3 & 37.7 & 60.9 & 23.6 & 64.6 & 43.6 & 63.7 & 47.1 & 61.1 & 15.4 & 62.6 & \graycell{}1.9 \\

        & \textbf{CodeT5} & 60.4 & 6.4 & 62.0 & 11.9 & 65.5 & \graycell{}4.3 & 67.3 & 11.6 & 58.6 & 4.6 & 59.6 & 5.7 \\

        & \textbf{UniXcoder} & 65.6 & 68.1 & 61.7 & 28.7 & 64.9 & 62.5 & 65.4 & 68.4 & 63.2 & 29.1 & 65.6 & \graycell{}1.2 \\

        \midrule
        \midrule

        & & \textbf{MRR} & \textbf{ANR} & \textbf{MRR} & \textbf{ANR} & \textbf{MRR} & \textbf{ANR} & \textbf{MRR} & \textbf{ANR} & \textbf{MRR} & \textbf{ANR} & \textbf{MRR} & \textbf{ANR} \\

        \midrule

        \multirow{3}{*}{\makecell[l]{\textbf{Code}\\\textbf{Search}}} & \textbf{CodeBERT} & 81.6 & 27.2 & 80.8 & 34.3 & 80.9 & 33.6 & 81.7 & 35.7 & 80.9 & 37.1 & 79.1 & \graycell{}49.3 \\

        & \textbf{CodeT5} & 81.8 & 28.1 & 81.7 & 30.8 & 80.9 & 33.2 & 82.3 & 32.8 & 81.1 & 33.0 & 79.0 & \graycell{}46.9 \\

        & \textbf{UniXcoder} & 82.6 & 26.5 & 83.0 & 32.2 & 82.5 & 34.2 & 82.0 & 35.4 & 81.6 & 38.2 & 78.0 & \graycell{}49.9 \\

        \midrule
        \midrule

        & & \textbf{BLEU} & \textbf{ASR} & \textbf{BLEU} & \textbf{ASR} & \textbf{BLEU} & \textbf{ASR} & \textbf{BLEU} & \textbf{ASR} & \textbf{BLEU} & \textbf{ASR} & \textbf{BLEU} & \textbf{ASR} \\

        \midrule

        \multirow{7}{*}{\makecell[l]{\textbf{Code}\\\textbf{Summarization}}} & \textbf{CodeBERT} & 19.0 & 8.3 & 18.3 & 12.6 & 18.8 & 9.0 & 19.3 & 7.35 & 18.2 & 6.2 & 18.3 & \graycell{}2.2 \\

        & \textbf{CodeT5} & 20.3 & 16.0 & 19.7 & 17.2 & 20.0 & 9.4 & 20.2 & 18.3 & 19.8 & 8.4 & 19.0 & \graycell{}6.3 \\

        & \textbf{UniXcoder} & 20.1 & 3.0 & 19.6 & 5.4 & 19.8 & 7.8 & 20.0 & 6.7 & 19.6 & 3.6 & 19.4 & \graycell{}2.0 \\

        \cmidrule(ll){2-14}

        & & \textbf{\majorR{METEOR}} & \textbf{ASR} & \textbf{\majorR{METEOR}} & \textbf{ASR} & \textbf{\majorR{METEOR}} & \textbf{ASR} & \textbf{\majorR{METEOR}} & \textbf{ASR} & \textbf{\majorR{METEOR}} & \textbf{ASR} & \textbf{\majorR{METEOR}} & \textbf{ASR} \\

        \cmidrule(ll){2-14}

        & \textbf{CodeBERT} & \majorR{27.7} & 8.3 & \majorR{25.9} & 12.6 & \majorR{26.7} & 9.0 & \majorR{29.0} & 7.35 & \majorR{25.7} & 6.2 & \majorR{27.1} & \graycell{}2.2 \\

        & \textbf{CodeT5} & \majorR{31.1} & 16.0 & \majorR{31.3} & 17.2 & \majorR{31.1} & 9.4 & \majorR{30.5} & 18.3 & \majorR{28.3} & 8.4 & \majorR{27.7} & \graycell{}6.3 \\

        & \textbf{UniXcoder} & \majorR{30.4} & 3.0 & \majorR{28.5} & 5.4 & \majorR{28.2} & 7.8 & \majorR{30.2} & 6.7 & \majorR{28.5} & 3.6 & \majorR{27.9} & \graycell{}2.0 \\

        \midrule
        \midrule

        & & \textbf{EM} & \textbf{ASR} & \textbf{EM} & \textbf{ASR} & \textbf{EM} & \textbf{ASR} & \textbf{EM} & \textbf{ASR} & \textbf{EM} & \textbf{ASR} & \textbf{EM} & \textbf{ASR} \\

        \midrule

        \multirow{3}{*}{\makecell[l]{\textbf{Code}\\\textbf{Repair}}} & \textbf{CodeBERT} & 15.2 & 5.9 & 12.8 & 2.7 & 14.6 & 2.8 & 14.4 & 3.3 & 14.5 & 2.8 & 14.6 & \graycell{}2.7 \\

        & \textbf{CodeT5} & 14.1 & 3.1 & 12.3 & 1.5 & 17.8 & 2.3 & 14.3 & 1.6 & 13.9 & 1.6 & 13.2 & \graycell{}1.5 \\

        & \textbf{UniXcoder} & 18.5 & 2.3 & 16.6 & 2.2 & 18.2 & 2.0 & 18.9 & 3.9 & 17.3 & 2.0 & 18.2 & \graycell{}0.7 \\
        
        \bottomrule
    \end{tabular}
    \begin{tablenotes}
       \item $^{*}$ The best defense results are highlighted in \hlgray{gray}.
    \end{tablenotes}
    \end{threeparttable}
    \vspace{-4mm}
\end{table*}

\noindent\textbf{Experimental Setup.}
\textit{Pre-training backdoor defenses} aim to prevent models from being implanted with backdoors by detecting and filtering poisoned samples from the training data before model training.
Activation Clustering (AC) and KillBadCode are effective in detecting samples with injected backdoor triggers.
\textit{AC}~\cite{2019-activation-clustering} clusters input representations via $K$-means and identifies the smaller cluster (below a threshold) as poisoned samples.
\textit{KillBadCode}~\cite{2025-KillBadCode} uses a $n$-gram model to detect poisoned code by identifying tokens whose removal improves perplexity, and removes samples containing such triggers.

\textit{In-training backdoor defenses} focus on preventing the insertion of backdoors during the training phase by actively monitoring and mitigating potential poisoning attempts.
DeCE~\cite{2024-DeCE} defends against backdoor attacks by using deceptive distributions and label smoothing to limit gradients, preventing overfitting to triggers during training.

\textit{Post-training backdoor defenses} are applied after the model has already been implanted with backdoors.
They can be further divided into backdoor elimination defenses and input detection defenses~\cite{2025-EliBadCode, 2024-CodePurify}.
Backdoor elimination defenses aim to remove backdoors from the model at the source, \majorD{such as EliBadCode~\cite{2025-EliBadCode}, which is discussed in detail in Section~\ref{subsec:reverse_engineering_technique}.}\majorR{for example through unlearning-based mitigation strategies~\cite{2025-EliBadCode, 2022-DBS}.}
\minorD{In contrast}\minorR{Notably}, input detection defenses focus on detecting anomalous (trigger-injected) inputs to prevent backdoors from being activated in the model. CodePurify~\cite{2024-CodePurify} detects and removes triggers in poisoned code using entropy-based scoring and a masked language model for purification.

\textit{Defense Settings.} 
For each defense method, we follow their original configurations, with detailed settings available in our repository~\cite{NatBackdoor}.
We evaluate the performance of \abbr{}s defended by these methods using test samples and use inverted triggers from RQ1 (Section~\ref{subsec:rq1}) to assess the effectiveness of these defenses in mitigating natural backdoor threats.

\noindent\textbf{Experimental Results.}
Table~\ref{tab:rq4} presents the effectiveness of different backdoor defense methods in mitigating natural backdoor vulnerabilities in \abbr{}s. The ``Undefended'' column represents \abbr{}s without any applied defenses, reflecting the original model utility together with the ASR/ANR of natural backdoor triggers.
\majorR{ Overall, most existing defenses designed for injected backdoors do not consistently mitigate natural backdoor threats across tasks and models. Although AC, KillBadCode, DeCE, and CodePurify reduce ASR or improve ANR in some individual settings, their effectiveness is unstable and they fail to provide consistent mitigation across different tasks.}
\majorD{It can be observed that AC, KillBadCode and CodePurify are ineffective in mitigating natural backdoor threats across different tasks, as they fail to reduce ASR or improve ANR.}
For instance, in the defect detection task, \majorD{the average ASR of AC, KillBadCode and CodePurify remains 21.39\%, 36.81\% and 24.56\%, respectively.}\majorR{AC reduces the ASR of CodeBERT from 37.7\% to 23.6\%, and CodePurify reduces it further to 15.4\%. However, these methods still leave relatively high ASR values for other models in the same task, and similar instability is observed across code search, code summarization, and code repair.}
This may be due to the fact that natural backdoor vulnerabilities are not intentionally implanted and lack distinct anomalous features, making them challenging for these defense methods to detect and mitigate.
\majorR{In addition, we observe that these defense methods may even increase the ASR of natural backdoor triggers. For example, in the defect detection task, DeCE increases the ASR of CodeBERT from 37.7\% to 47.1\%, and that of CodeT5 from 6.4\% to 11.6\%. These results suggest that defense methods designed for injected backdoors may not directly generalize to natural backdoors, because the trigger patterns of natural backdoors are more deeply entangled with normal data distributions.}\majorD{ Unexpectedly, DeCE not only fails to mitigate natural backdoor vulnerabilities but even amplifies their impact.
For example, in the defect detection task, DeCE increases the ASR from 37.7\% to 47.1\%.
This could be attributed to DeCE’s focus on reducing overfitting to ``anomalous'' patterns, which inadvertently causes the model to overfit patterns that appear ``normal'' but actually trigger natural backdoors, exacerbating their effect.}
In contrast, \majorD{EliBadCode}\majorR{the unlearning-based defense} demonstrates effectiveness in mitigating natural backdoor vulnerabilities across different tasks.
Specifically, \majorD{EliBadCode}\majorR{it} reduces the average ASR to 3.0\%, 3.5\%, and 1.7\% in the defect detection, code summarization, and code repair tasks, respectively, while increasing the average ANR to 48.7\% in the code search task.
\majorD{This effectiveness can be attributed to the model unlearning approach in EliBadCode. Model unlearning}\majorR{This effectiveness may stem from the fact that it} constructs a small set of ``inverse'' samples (i.e., samples that leverage the reversed association between trigger samples and target labels) and optimizes the model using these samples, thereby disrupting the erroneous mapping between natural backdoor triggers and target labels.

\majorR{In addition, we observe that these defense methods do not cause severe utility degradation to the models. For example, for the unlearning-based defense, the defended models maintain ACC values comparable to those of the undefended models in defect detection (e.g., CodeBERT: 61.3\% vs. 62.6\%; CodeT5: 60.4\% vs. 59.6\%; UniXcoder: 65.6\% vs. 65.6\%). In code search, the defended models show only limited fluctuations in MRR and largely preserve retrieval capability close to the undefended setting. In code summarization, the defended models retain similar BLEU scores and remain competitive under the semantic-aware metric METEOR. In code repair, the EM scores also remain close to those of the undefended setting. These results suggest that, although the existing defense methods differ in their security improvements, they do not substantially impair the normal task capabilities of the models overall. Among them, the unlearning-based defense is able to reduce the effectiveness of natural backdoor attacks while largely preserving model utility, thus achieving a relatively favorable security-utility trade-off.}

\summary[Answer to RQ5]{Experimental results indicate that most existing backdoor defense methods for \abbr{}s are ineffective in mitigating the impact of natural backdoor vulnerabilities.
In contrast, \majorD{EliBadCode}\majorR{unlearning-based defense} shows promise as an effective defense technique for mitigating the risks associated with these vulnerabilities.}

\subsection{RQ6: An Enhanced Natural Backdoor Detection Method}
\label{subsec:rq6}

\begin{algorithm}[!t]
    \caption{Inversion of Natural Backdoor Triggers}
    \scriptsize
    \label{alg:scannbt}
    \raggedright
    \begin{tabular}{rllll}
        \hline
        \textsc{Input}: 
        & $X$, $Y$ & \; & clean samples, target label & \\
        & $f_{\theta}$, $V$ & \; & clean \abbr{}, trigger vocabulary & \\
        & $R$, $I$ & \; & number iterations, maximum updates per round & \\
        & $n$, $\alpha$ & \; & trigger length, patience threshold for ASR stagnation & \\
        \textsc{Output}: 
        & $\mathcal{T}$ & \; & set of natural backdoor triggers & \\
        \hline
    \end{tabular}
    \begin{algorithmic}[1]
        \Function{TriggerInversion}{$S, y'$}
            \State $T \gets \emptyset$ \hfill\Comment{\textcolor{gray}{set of effective triggers}}
            \State $U \gets \emptyset$ \hfill\Comment{\textcolor{gray}{set of fixed trigger tokens}}

            \For{$r = 1$ to $R$}
                \State $t \gets \textsc{RandomInitTrigger}(V \setminus U, n)$ \hfill\Comment{\textcolor{gray}{initialize a trigger of $n$ tokens sampled from the allowed vocabulary}}

                \State $a_{\mathrm{best}} \gets 0,\; t_{\mathrm{best}} \gets t,\; m \gets 0$ \hfill\Comment{\textcolor{gray}{best ASR, best trigger, stagnation counter}}

                \For{$i = 1$ to $I$}
                    \State $t \gets \Call{TriggerInversionStep}{f_{\theta}, S, y', t}$ \hfill\Comment{\textcolor{gray}{perform one inversion update step}}

                    \State $a \gets \Call{ComputeASR}{f_{\theta}, S, y', t}$ \hfill\Comment{\textcolor{gray}{compute ASR}}

                    \If{$a > a_{\mathrm{best}}$}
                        \State $a_{\mathrm{best}} \gets a, t_{\mathrm{best}} \gets t$
                        \State $m \gets 0$ \hfill\Comment{\textcolor{gray}{reset stagnation counter}}
                    \Else
                        \State $m \gets m + 1$ \hfill\Comment{\textcolor{gray}{no improvement}}
                    \EndIf

                    \If{$m = \mathrm{patience}$}
                        \State \textbf{break} \hfill\Comment{\textcolor{gray}{early stopping due to stagnation}}
                    \EndIf
                \EndFor

            \State $T \gets T \cup \{t_{\mathrm{best}}\}$ \hfill\Comment{\textcolor{gray}{record the most effective trigger for this round}}

            \State $U \gets U \cup \Call{Tokenizer}{t_{\mathrm{best}}}$ \hfill\Comment{\textcolor{gray}{update fixed trigger tokens}}

            \EndFor

            \State \Return $T$
        \EndFunction
        \\

        \State $\mathcal{T} \gets \emptyset$  \hfill\Comment{\textcolor{gray}{store inverted triggers}}
        \For{each label $y'$ \textbf{in} $Y$}
            \State $S \gets$ get code snippets in $X$ according to $y'$
            \State $T \gets $ \Call{TriggerInversion}{$S$, $y'$}
            \State $\mathcal{T}[y'] \gets T$
        \EndFor
        \State \textbf{Output} $\mathcal{T}$
    \end{algorithmic}
\end{algorithm}

% \begin{table}[t]
\begin{table}[htbp]
    \centering
    \scriptsize
    \tabcolsep=1.5pt
    \caption{Performance of \ours{} and EliBadCode in exposing natural backdoor vulnerabilities in CodeBERT. \majorR{\ours{}-NR denotes the variant of ScanNBT without reinitialization.}}
    \vspace{-1mm}
    \label{tab:rq5}
    \begin{threeparttable}
    \begin{tabular}{ccccccccccccc}
        \toprule

        \textbf{ID} & \multicolumn{4}{c}{\textbf{EliBadCode}} & \multicolumn{4}{c}{\majorR{\textbf{\ours{}-NR}}} & \multicolumn{4}{c}{\textbf{\ours{}}} \\

        \midrule
        \midrule
        
        \multicolumn{13}{c}{\textbf{Defect Detection}} \\
        \midrule
        & \textbf{ASR} & \textbf{D-1} & \textbf{D-2} & \majorR{\textbf{Time}} & \textbf{ASR} & \textbf{D-1} & \textbf{D-2} & \majorR{\textbf{Time}} & \textbf{ASR} & \textbf{D-1} & \textbf{D-2} & \majorR{\textbf{Time}} \\

        \midrule

        $\boldsymbol{D_1}$ & \graycell{}9.26 & - & - & \majorR{15m31s} & 8.26 & 0.80 & 1.00 & \majorR{3m48s} & 8.18 & \graycell{}0.92 & \graycell{}1.00 & \majorR{15m19s} \\
        $\boldsymbol{D_2}$ & 32.46 & 0.21 & 0.41 & \majorR{15m29s} & 34.25 & 0.37 & 0.64 & \majorR{3m52s} & \graycell{}35.88 & \graycell{}0.93 & \graycell{}1.00 & \majorR{15m23s} \\

        \midrule
        \midrule

        \multicolumn{13}{c}{\textbf{Code Search}} \\
        \midrule
        & \textbf{ANR} & \textbf{D-1} & \textbf{D-2} & \textbf{Time} & \textbf{ANR} & \textbf{D-1} & \textbf{D-2} & \textbf{Time} & \textbf{ANR} & \textbf{D-1} & \textbf{D-2} & \textbf{Time} \\

        \midrule

        $\boldsymbol{S_1}$ & 37.76 & 0.16 & 0.44 & \majorR{30m07s} & 35.17 & 0.31 & 0.55 & \majorR{3m09s} & \graycell{}34.13 & \graycell{}0.80 & \graycell{}1.00 & \majorR{33m22s} \\
        $\boldsymbol{S_2}$ & 34.24 & 0.17 & 0.40 & \majorR{13m22s} & 33.78 & 0.60 & 0.88 & \majorR{4m48s} & \graycell{}32.28 & \graycell{}0.67 & \graycell{}1.00 & \majorR{13m21s} \\

        \midrule
        \midrule

        \multicolumn{13}{c}{\textbf{Code Summarization}} \\
        \midrule
        & \textbf{ASR} & \textbf{D-1} & \textbf{D-2} & \majorR{\textbf{Time}} & \textbf{ASR} & \textbf{D-1} & \textbf{D-2} & \majorR{\textbf{Time}} & \textbf{ASR} & \textbf{D-1} & \textbf{D-2} & \majorR{\textbf{Time}} \\

        \midrule

        $\boldsymbol{M_1}$ & \graycell{}7.00 & 0.37 & 0.67 & \majorR{3h54m22s} & 6.12 & 0.40 & 0.88 & \majorR{41m0s} & 6.71 & \graycell{}0.48 & \graycell{}0.90 & \majorR{3h51m54s} \\
        $\boldsymbol{M_2}$ & \graycell{}11.91 & 0.53 & 0.75 & \majorR{4h57m43s} & 10.98 & 0.40 & 0.56 & \majorR{45m10s} & 10.36 & \graycell{}0.72 & \graycell{}1.00 & \majorR{5h04m25s} \\
        $\boldsymbol{M_3}$ & 6.72 & - & - & \majorR{5h14m32s} & 6.99 & 0.43 & 0.75 & \majorR{49m19s} & \graycell{}10.01 & \graycell{}0.44 & \graycell{}0.80 & \majorR{5h05m31s} \\
        $\boldsymbol{M_4}$ & 1.73 & 0.28 & 0.35 & \majorR{5h02m55s} & 1.27 & 0.60 & 1.00 & \majorR{28m05s} & \graycell{}2.49 & \graycell{}0.67 & \graycell{}1.00 & \majorR{4h45m38s} \\

        \midrule
        \midrule

        \multicolumn{13}{c}{\textbf{Code Repair}} \\
        \midrule
        & \textbf{ASR} & \textbf{D-1} & \textbf{D-2} & \majorR{\textbf{Time}} & \textbf{ASR} & \textbf{D-1} & \textbf{D-2} & \majorR{\textbf{Time}} & \textbf{ASR} & \textbf{D-1} & \textbf{D-2} & \majorR{\textbf{Time}} \\

        \midrule

        $\boldsymbol{R_1}$ & \graycell{}12.21 & 0.19 & 0.45 & \majorR{5h01m03s} & 9.58 & 0.40 & 0.45 & \majorR{39m39s} & 11.61 & \graycell{}0.91 & \graycell{}1.00 & \majorR{5h09m28s} \\
        $\boldsymbol{R_2}$ & 5.97 & 0.19 & 0.43 & \majorR{5h11m10s} & 6.41 & 0.44 & 0.65 & \majorR{35m29s} & \graycell{}6.77 & \graycell{}1.00 & \graycell{}1.00 & \majorR{5h03m55s} \\
        $\boldsymbol{R_3}$ & \graycell{}1.92 & 0.16 & 0.33 & \majorR{5h25m53s} & 1.91 & - & - & \majorR{38m07s} & 1.83 & \graycell{}0.91 & \graycell{}1.00 & \majorR{5h50m22s} \\
        $\boldsymbol{R_4}$ & \graycell{}5.34 & 0.18 & 0.42 & \majorR{5h07m50s} & 4.95 & 0.36 & 0.64 & \majorR{30m58s} & 4.67 & \graycell{}0.88 & \graycell{}1.00 & \majorR{5h50m31s} \\

        \bottomrule
    \end{tabular}
    \begin{tablenotes}
        \item $^*$ D-1: Distinct-1; D-2: Distinct-2.
        \item $^{**}$ The best ASR/ANR and Distinct-1/2 results are highlighted in \hlgray{gray}.
        \item $^{***}$ ``-'' indicates that only a single effective inverted trigger is available, making the computation of Distinct-n infeasible.
    \end{tablenotes}
    \end{threeparttable}
    \vspace{-4mm}
\end{table}

\noindent\textbf{Design.}
In RQ5 (Section~\ref{subsec:rq5}), we show that \majorD{EliBadCode}\majorR{the unlearning-based defense} effectively mitigates natural backdoor vulnerabilities in \abbr{}s\majorD{ through model unlearning}. However, model unlearning can only remove detected backdoor vulnerabilities, highlighting the importance of comprehensively detecting natural backdoor vulnerabilities in \abbr{}s to enhance security.

EliBadCode is designed to detect maliciously implanted backdoors and construct optimal trigger tokens through continuous optimization.
However, this optimization strategy \majorR{focuses on a single locally optimal trigger,} limiting the diversity of the generated triggers\majorR{ and thus reducing} its ability to comprehensively expose potential natural backdoor vulnerabilities in \abbr{}s.
To address this limitation, we propose a novel \underline{\textbf{scan}}ning for \underline{\textbf{n}}atural \underline{\textbf{b}}ackdoor \underline{\textbf{t}}riggers method, \ours{}, which enhances the exposure of natural backdoor vulnerabilities in \abbr{}s by introducing trigger fixation and re-initialization. \majorR{The goal of \ours{} is not merely to converge to a single optimal trigger, but to progressively accumulate a diverse set of effective triggers across multiple search rounds.}

\begin{figure}[!t]
    \centering
    \includegraphics[width=\linewidth]{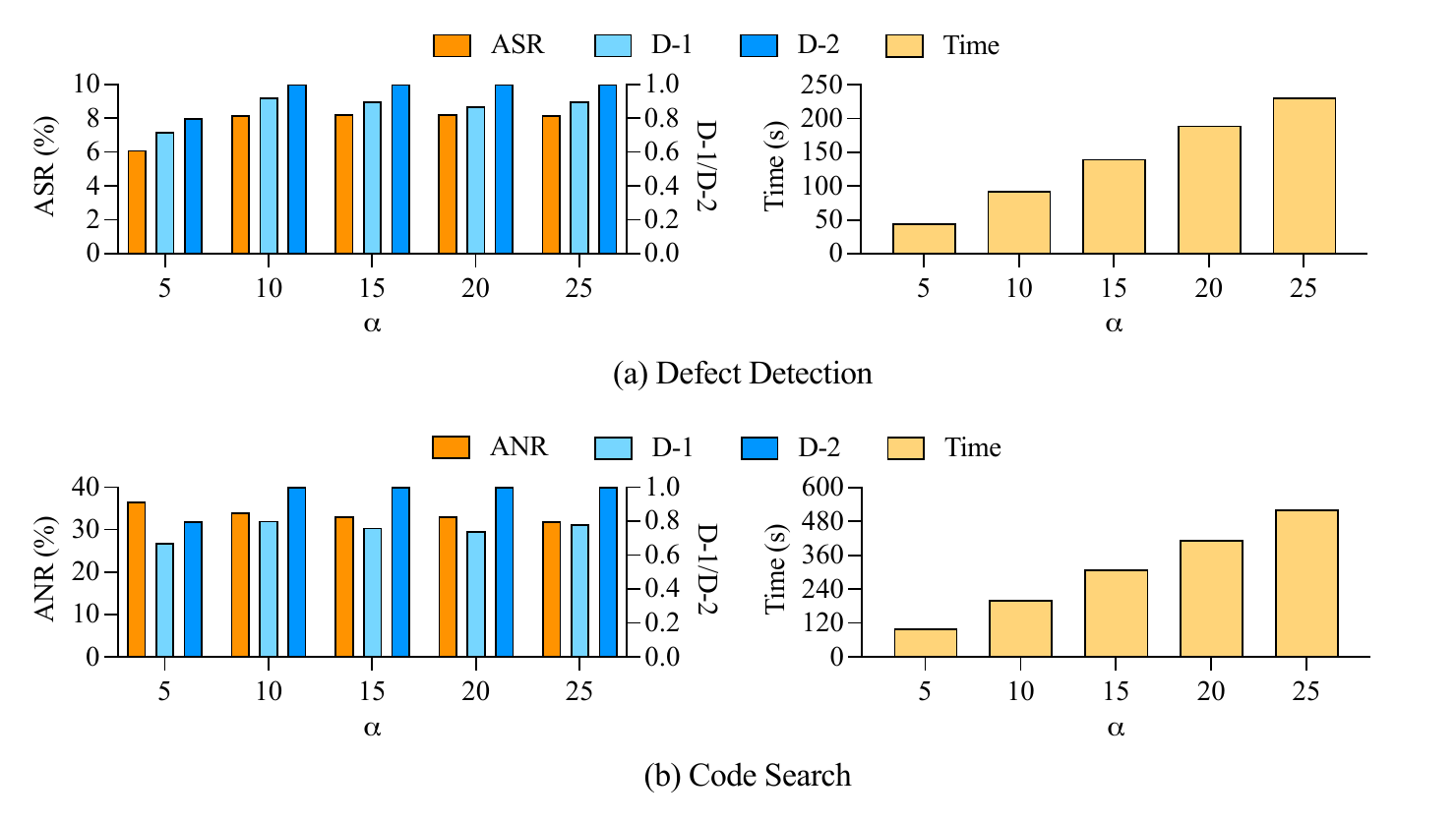}
    \vspace{-6mm}
    \caption{\minorR{Sensitivity results of the patience threshold $\alpha$ in \ours{} on CodeBERT for defect detection ($\mathcal{D}_1$) and code search ($\mathcal{S}_1$). Time denotes the average runtime per re-initialization round.}}
    \label{fig:a}
    \vspace{-4mm}
\end{figure}

Algorithm~\ref{alg:scannbt} illustrates the trigger inversion procedure of \ours{} in detail. Specifically, \ours{} first monitors the trend of ASR changes during the optimization process (lines 9-18). If ASR does not improve for $\alpha$ consecutive iterations, the trigger tokens corresponding to the epoch at which ASR reaches its highest value are fixed and recorded as effective trigger tokens (lines 10-12).
This fixation preserves valuable triggers identified within the current search space\majorR{, so that \ours{} retains the best trigger found in each round and gradually accumulates a set of effective triggers across rounds, rather than keeping only a single global optimum.}
The rationale behind this is that when ASR fails to improve for \majorR{$\alpha$} consecutive iterations, it may indicate that the trigger has reached a local optimum. The fixed trigger token combination, corresponding to the highest ASR, is considered an effective trigger capable of activating a natural backdoor. This process is similar to the early stopping mechanism\majorR{~\cite{2011-Scikit-learn, 2019-Auto-Keras}.}
\minorR{Figure~\ref{fig:a} presents the sensitivity results on CodeBERT for $\alpha \in \{5, 10, 15, 20, 25\}$ across defect detection ($\mathcal{D}_1$) and code search ($\mathcal{S}_1$). It can be observed that as $\alpha$ increases, ASR/ANR and trigger diversity (Distinct-1/2) gradually stabilize, while the average runtime per re-initialization round increases substantially. This suggests that $\alpha = 5$ leads to premature re-initialization before the current search space has been sufficiently explored, whereas $\alpha > 10$ substantially increases the runtime with no meaningful gain in attack effectiveness or trigger diversity. Therefore, we set $\alpha$ to 10 in \ours{}, as it achieves a favorable balance between exploration thoroughness and computational efficiency.}
Subsequently, \ours{} reinitializes the trigger tokens to explore new effective triggers in a different search space (line 5).
\majorR{Importantly, this re-initialization is memory-guided rather than a random restart over the full vocabulary.}
\majorD{It is worth noting that, in}\majorR{In} the next optimization round, the tokens being optimized exclude previously recorded effective trigger tokens (line 21).
\majorD{This prevents redundant searches of discovered triggers and encourages the optimization process to explore new potential triggers.}
\majorR{This prevents redundant searches of already discovered triggers and guides the optimization process toward regions of the trigger space that have not yet been explored, thereby improving trigger diversity and trigger-space coverage for natural backdoor exposure.}

\noindent\textbf{Experimental Setup.}
In the experiments, for \ours{}, we treat all inversion-fixed triggers (Algorithm~\ref{alg:scannbt}, line 20) as potential natural backdoor triggers.
For EliBadCode, we follow its original criterion and retain inversion-generated triggers whose ASR or ANR is within 10 percentage points of the optimal trigger.
\majorR{Additionally, we introduce a variant of \ours{}, denoted as \ours{}-NR (No Reinitialization). When ASR fails to improve for $\alpha$ consecutive iterations, this variant terminates the search without reinitialization and retains the trigger set identified up to that point, including the effective trigger selected by trigger fixation and those inversion-generated triggers whose ASR or ANR is within 10 percentage points of it. We do not introduce a ``No Fixation'' variant because, in \ours{}, trigger fixation also serves as the stagnation signal for reinitialization; removing it would eliminate the mechanism for starting a new search round and cause the method to degenerate into EliBadCode.}
We evaluate the attack effectiveness of the potential natural backdoor triggers using ASR and ANR, and assess their diversity using Distinct-1 and Distinct-2.

\noindent\textbf{Experimental Results.} 
Table~\ref{tab:rq5} shows the performance of \ours{}, \ours{}-NR, and EliBadCode in exposing natural backdoor vulnerabilities in CodeBERT across four tasks: defect detection, code search, code summarization, and code repair.

It can be observed that across all tasks, \ours{} consistently outperforms EliBadCode in terms of trigger diversity, as reflected by higher Distinct-1 and Distinct-2 scores. For example, in the defect detection task, EliBadCode achieves only 0.21/0.41 in Distinct-1/2, whereas \ours{} substantially improves these scores to 0.93/1.00.
\majorR{Similar gains can also be observed in code search, code summarization, and code repair. These results indicate that \ours{} is able to expose a more diverse set of effective natural backdoor triggers, rather than concentrating on only a few locally optimal trigger patterns.}
In addition to improving diversity, the triggers exposed by \ours{} also exhibit stronger attack effectiveness. Specifically, \ours{} achieves a higher average ASR in defect detection (22.03\% vs. 20.86\%) and code summarization (7.39\% vs. 6.84\%), and a lower average ANR in code search (33.21\% vs. 36.00\%) than EliBadCode. These improvements indicate that \ours{} is capable of generating not only more diverse but also more effective triggers.
Compared with \ours{}-NR, \ours{} further improves diversity in most settings, which shows the benefit of reinitialization beyond trigger fixation alone. For instance, in defect detection, \ours{} improves Distinct-1/2 from 0.80/1.00 and 0.37/0.64 under \ours{}-NR to 0.92/1.00 and 0.93/1.00. In code summarization and code repair, \ours{} also consistently yields higher Distinct scores in most cases. These results suggest that reinitialization helps the search escape the current trigger region and discover additional effective triggers, rather than merely repeating the same inversion process.

\majorR{In addition, we observe that the performance gains of \ours{} come with only acceptable additional computational cost. For example, in defect detection and code search, \ours{} typically requires about 15-33 minutes, while EliBadCode requires about 13-30 minutes. In \minorD{contrast}\minorR{comparison}, \ours{}-NR runs much faster because, once optimization stagnation is detected, it does not perform reinitialization or enter a new search round. These results suggest that iterative reinitialization introduces some extra runtime overhead but does not substantially increase the overall computational cost. At the same time, it significantly improves trigger diversity and, in most tasks, improves or preserves the attack effectiveness for exposing natural backdoor vulnerabilities.}

\summary[Answer to RQ6]{\ours{} exposes more diverse and effective triggers by introducing trigger fixation and re-initialization, thereby improving the comprehensive detection of natural backdoor vulnerabilities in \abbr{}s \majorR{with acceptable additional runtime cost.}}

\section{Discussion}
\label{sec:discussion}

\subsection{Implications of Natural Backdoor Vulnerabilities in Practical Scenarios}

In practical scenarios, natural backdoor vulnerabilities in CodeLMs can pose significant security risks. As demonstrated in RQ3, when multiple \abbr{}s share similar fine-tuning datasets, model architectures, or high-level learned representations, natural backdoor triggers may transfer across models. This cross-model transferability implies that once a natural backdoor exists in one model, its effects may propagate across its related models.
For adversaries, such transferability substantially lowers the barrier to practical exploitation. Even without direct access to the target \abbr{}, an attacker can approximate the victim model’s behavior using model fingerprinting techniques~\cite{2025-LLMmap, 2025-SeedPrints} or membership inference attacks~\cite{2024-Membership-Inference-Attacks-Against-In-Context-Learning, 2025-Towards-Label-Only-Membership-Inference-Attack}. By identifying natural trigger patterns on these surrogate models, the attacker can construct inputs that reliably activate the corresponding vulnerabilities in the target system.

Natural backdoor vulnerabilities can also pose risks in seemingly benign development scenarios. For example, a developer may inadvertently enter code fragments that resemble a trigger. A minor typographical error (such as a misspelled variable name or a slightly modified identifier) may unexpectedly match a natural trigger pattern. As demonstrated in Figure~\ref{fig:case}, such subtle changes can significantly alter the predictions of a CodeLM. Given these risks, we underscore the importance of proactively identifying and mitigating natural backdoor triggers in \abbr{}s.

\subsection{Natural Backdoor Triggers in Code Structures}
Currently, structural backdoors in code have primarily been explored in the form of dead code~\cite{2022-Backdoors-in-Neural-Models-of-Source-Code, 2022-you-see-what-I-want-you-to-see}, where injected but unused code segments serve as backdoor triggers. 
Therefore, natural backdoor vulnerabilities in \abbr{}s may not be restricted to variable or method names; they could arise from structural properties such as code indentation patterns, control flow structures (e.g., for, while loops), or code formatting styles. Such structural features may also serve as natural backdoor triggers.
\majorD{However, existing  trigger inversion techniques mainly rely on token-level analysis to invert potential backdoor triggers, making them less effective in identifying and reconstructing structural backdoor triggers.
This limitation arises because
structural triggers are inherently more challenging to invert than those based on variable or method names.  
On one hand, code structure is deeply intertwined with program execution logic, meaning forced modifications could inadvertently alter its semantics. On the other hand, reconstructed structural triggers may not always retain syntactic correctness, further complicating the inversion process.}
\majorR{However, current gradient-based trigger inversion methods can effectively invert only token-level backdoor triggers. In contrast, code structures (e.g., indentation patterns and control-flow constructs) are constrained by hard requirements such as parsing, compilation, type checking, and syntactic well-formedness. These structural properties cannot be naturally parameterized as differentiable variables for gradient-based optimization in existing methods, and thus such methods cannot reliably invert structure-based triggers.}
Following previous work, our study focuses on \majorR{inverting} natural backdoor triggers based on variable- or method-name tokens, as they align with the capabilities of current trigger inversion techniques while still providing meaningful insights into the security of \abbr{}s.

Given that structural features are pervasive in code and may lead to more stealthy natural backdoor triggers, existing work remains limited in its ability to reliably invert and detect structure-based triggers. Therefore, developing structure-aware trigger inversion and detection methods is an important direction for future work. \minorR{Because code structure is fundamentally topological in nature, such methods should move beyond discrete token sequences and instead operate on graph-based code representations. For instance, tools such as Joern can parse source code into Code Property Graphs (CPGs)~\cite{2014-Modeling-and-Discovering-Vulnerabilities-with-Code-Property-Graphs}, which unify syntactic, control-flow, and data-flow information into a single graph structure. Building on this, future approaches may leverage Graph Neural Networks (GNNs) or graph-level explainability techniques to identify connected subgraphs or vulnerability paths that exhibit anomalous label correlations, thereby enabling effective detection of structural natural backdoors while preserving topological continuity and syntactic well-formedness.}

\subsection{\minorR{Limitations of Unlearning-Based Defense Against Zero-Day Natural Backdoors}}
\minorR{Despite its effectiveness, the unlearning-based defense has an inherent limitation regarding generalizability to un-scanned natural backdoors. Specifically, the defense operates by constructing inverse samples based on the triggers identified through the trigger inversion pipeline, and then fine-tuning the model on these inverse samples to disrupt the erroneous mapping between the identified triggers and target labels. Consequently, its protection is fundamentally bounded by the coverage of the trigger inversion phase: natural backdoor vulnerabilities that were not discovered during trigger inversion (i.e., zero-day natural backdoors) cannot be directly addressed by this defense. As demonstrated in RQ6, ScanNBT is able to expose a more diverse set of effective triggers beyond those found by EliBadCode, which suggests that the trigger set used to construct inverse samples in RQ5 may not be exhaustive. In practice, a CodeLM may harbor multiple natural backdoor vulnerabilities simultaneously, and an attacker who discovers a vulnerability overlooked by the trigger inversion pipeline could still exploit the defended model with near-original ASR.
This reveals a core limitation of unlearning-based defenses: their effectiveness is inherently bounded by the completeness of trigger discovery, yet trigger inversion is fundamentally incomplete. Future work could explore complementary strategies that do not rely on explicit trigger identification, so as to provide broader protection against both scanned and zero-day natural backdoor vulnerabilities.}

\subsection{\majorR{Natural Backdoors as a Manifestation of Shortcut Learning}}
\majorR{Our findings in RQ4 suggest that natural backdoors may be closely related to shortcut learning in deep neural networks. Shortcut learning refers to the phenomenon that models tend to rely on simple but spuriously predictive statistical cues, rather than learning robust high-level semantic information~\cite{2020-Shortcut-Learning-in-Deep-Neural-Networks, 2020-Robustness-to-Spurious-Correlations-using-Pre-trained-Language-Models, 2024-Spurious-Correlations-in-Machine-Learning-A-Survey}. In our setting, the identified high-Z-score tokens exhibit strong statistical correlations with the target labels, suggesting that the model may exploit these tokens as ``shortcuts'' for prediction.
Moreover, as shown in Figure~\ref{fig:causal_intervention}, removing the high-Z-score biased token from an inverted trigger leads to a much larger reduction in attack effectiveness than removing non-biased tokens. This indicates that these tokens are not merely correlated with the target labels, but play a causal role in activating the observed natural backdoor behaviors. In other words, the model appears to over-rely on these statistically salient tokens, rather than on more semantically grounded program features, which is consistent with the phenomenon of shortcut learning, where models rely on superficial cues instead of robust semantic features.
From this perspective, natural backdoors can be viewed as a manifestation of shortcut learning under biased data distributions. Unlike injected backdoors, where the association between triggers and target labels is deliberately implanted by attackers, natural backdoors emerge when models unintentionally learn spurious shortcut features from clean but biased data~\cite{2022-Backdoor-Vulnerabilities-in-Normally-Trained-Deep-Learning-Models, 2023-PELICAN}. Once such features are learned, they may function as trigger-like signals that induce abnormal predictions or unsafe retrieval behaviors. Therefore, mitigating natural backdoors may also require reducing shortcut learning, for example by alleviating dataset bias and encouraging models to rely more on semantically meaningful program features.}

\subsection{\majorR{Natural Backdoors and Legitimate Causal Features}}
\majorR{A related question is how to distinguish natural backdoor-related features from legitimate causal features that are semantically justified by the task itself. We do not define legitimate causal features as natural backdoors. For example, in vulnerability-related tasks, APIs or functions such as \texttt{strcpy} may reasonably contribute to prediction because they are directly associated with unsafe memory operations. Therefore, such features should not be interpreted as natural backdoors merely because they are highly predictive.
Instead, we argue that natural backdoor risk arises from the model's abnormal reliance on a feature, rather than from the feature itself. Under biased data distributions, even semantically justified features may be over-amplified and used by the model in a shortcut-like or disproportionately target-biased manner. In such cases, the feature may exhibit natural-backdoor-like behavior, not because the feature itself is illegitimate, but because the model relies on it far more strongly than broader semantic evidence supports. From this perspective, our method is intended to identify candidate trigger-like vulnerabilities associated with biased or shortcut-like model behavior, rather than to label all influential features as natural backdoors.
We further note that the boundary between legitimate causal features and natural backdoor-related features is not always perfectly clear in practice. Some features may simultaneously encode semantically grounded task information and biased statistical associations learned from the training data. Therefore, our analysis should be interpreted as a vulnerability-oriented diagnosis: it highlights features whose predictive effects appear abnormally strong and may induce disproportionate prediction shifts, and thus deserve further inspection.}

\section{Threats to Validity}
Our empirical study may contain several threats to external and internal validity, which we have attempted to mitigate.

\noindent\textbf{Threats to Internal Validity.}
A threat to internal validity lies in the applicability of natural backdoor triggers. We focus on triggers in code inputs to reveal natural backdoor vulnerabilities in \abbr{}s. However, specific patterns in other non-code input forms (e.g., comments or natural language queries) may also trigger natural backdoor vulnerabilities in \abbr{}s. Considering that the inverted triggers are composed of tokens, and triggers in other non-code inputs are also token-based, we have high confidence in the generalization capability of natural backdoor triggers across different input scenarios. In future work, we will further investigate how non-code inputs may induce natural backdoor vulnerabilities in \abbr{}s and their potential security implications.

\minorR{In RQ1, we rely solely on EliBadCode to identify natural backdoor triggers, which are then reused in the defense evaluation of RQ5. This raises a potential concern that the superior performance of the unlearning-based defense may be partly influenced by the fact that the evaluated triggers were discovered by the same technique. Nevertheless, this concern is partially mitigated by several observations. First, natural backdoor triggers are prior vulnerabilities arising from training-data bias, rather than artifacts introduced by EliBadCode itself. Second, the unlearning-based defense consistently outperforms methods unrelated to trigger inversion (e.g., AC, KillBadCode, and DeCE), suggesting that its advantage stems more from the unlearning mechanism than from the trigger source. Third, triggers inverted by \ours{} and \ours{}-NR under different exploration strategies also achieve high ASR, indicating that the exposed vulnerabilities are not unique to EliBadCode's search behavior. Incorporating more diverse trigger inversion techniques in future work would further strengthen the reliability of these conclusions.}

\noindent\textbf{Threats to External Validity.}
A threat to external validity lies in the generalizability of our findings. We investigate natural backdoor vulnerabilities in a diverse set of \abbr{}s, including CodeBERT, CodeT5, UniXcoder, StarCoder, DeepSeek-Coder, and GPT-3.5, spanning multiple programming languages and code intelligence tasks.
\majorR{However, due to computational resource constraints, not all analyses can be conducted uniformly across all models. In particular, several fine-grained analyses mainly focus on CodeBERT, CodeT5, and UniXcoder as the primary analysis subjects. Compared with recent large-scale models, these models are relatively small, which may limit the generalizability of those detailed observations.}
\majorD{However}\majorR{In addition}, it remains unclear whether the conclusions of our experiment can be maintained in \majorR{a broader range of} other \abbr{}s.
To mitigate this threat, we systematically examine \abbr{}s with varying architectures and parameter scales, encompassing both pre-trained and large-scale models. Furthermore, our experiments include both fine-tuned and off-the-shelf models, along with black-box and white-box scenarios, to provide a more comprehensive evaluation of natural backdoor vulnerabilities in \abbr{}s.

\majorR{Another threat to external validity concerns the choice of victim and target labels across different tasks. In the code summarization task, flipping antonym keywords (e.g., \texttt{open}$\rightarrow$\texttt{close} or \texttt{read}$\rightarrow$\texttt{write}) may be more readily viewed as a semantic or quality-level error rather than a directly exploitable security vulnerability. However, in real-world development, such semantic manipulation can still mislead developers during code comprehension, review, or reuse, causing them to misinterpret a function’s behavior and potentially make incorrect or unsafe integration decisions. In contrast, in the code repair task, we flip key symbols/constants with clear functional semantics (e.g., \texttt{==}$\rightarrow$\texttt{!=}, \texttt{true}$\rightarrow$\texttt{false}), which directly changes program logic and conditional checks and more explicitly reflects potential security impact. Therefore, the impact of natural backdoors may vary across downstream scenarios; our results should primarily be interpreted as demonstrating the existence and transferability of such vulnerabilities, while their exploit severity depends on the specific application setting.}

\section{\majorR{Related Work}}
\label{sec:related_work}

\subsection{\majorR{Injected Backdoor Attacks in \abbr{}s}}
\majorR{Recent studies show that \abbr{}s are vulnerable to injected backdoor attacks~\cite{2024-Security-of-Language-Models-for-Code}. 
A backdoored \abbr{} may preserve normal performance on clean inputs while returning defective, vulnerable, or otherwise unsafe code when the trigger is present, thereby posing security risks to downstream software systems~\cite{2025-KillBadCode, 2025-EliBadCode}.
Early studies typically use dead-code insertion as the trigger, including both fixed triggers and grammar-based triggers, and demonstrate that even poisoning only a small portion of the training data can successfully implant backdoors into code models~\cite{2022-Backdoors-in-Neural-Models-of-Source-Code, 2022-you-see-what-I-want-you-to-see}. 
For example, Ramakrishnan and Albarghouthi~\cite{2022-Backdoors-in-Neural-Models-of-Source-Code} use fixed or grammar-based dead-code snippets as triggers and validate the effectiveness of such attacks on code summarization and method name prediction tasks. 
Wan et al.~\cite{2022-you-see-what-I-want-you-to-see} further extend this idea to code search, showing that poisoned samples containing dead-code triggers can significantly promote malicious code under specific target queries.
Subsequent studies replace dead-code triggers with identifier-based triggers, such as variable/function renaming or name extension, to improve stealthiness, since such triggers are more difficult for developers and static analysis tools to detect~\cite{2023-BADCODE, 2025-Hidden-Backdoor-Attack, 2024-Stealthy-Backdoor-Attack-for-Code-Models}. 
For example, BadCode~\cite{2023-BADCODE} and HiBadCode~\cite{2025-Hidden-Backdoor-Attack} construct more stealthy triggers by appending specific tokens to existing function or variable names, and significantly outperform fixed and grammar-based triggers in code search. 
AFRAIDOOR~\cite{2024-Stealthy-Backdoor-Attack-for-Code-Models} further combines adversarial perturbations with identifier renaming to adaptively generate different triggers for different samples, thereby improving both stealthiness and evasion capability.
In addition, Li et al.~\cite{2023-multi-target-backdoor-attacks} further investigate model poisoning as another form of injected backdoor attack, showing that poisoned pre-trained models may retain their backdoor behaviors even after downstream fine-tuning on clean data.}

\majorR{Although these studies demonstrate that \abbr{}s are vulnerable to trigger-based manipulation, they generally assume an active adversary who deliberately implants the trigger-target association through poisoned data or poisoned models. 
In contrast, this paper studies natural backdoors in \abbr{}s, namely vulnerabilities that arise in normally trained models without malicious data poisoning or adversarial manipulation.}

\subsection{\majorR{Natural Backdoors in Normally Trained Models}}

\majorR{Beyond injected backdoors, normally trained language models may also exhibit natural backdoors, namely trigger-like vulnerabilities that emerge unintentionally from benign training data rather than adversarial poisoning~\cite{2022-Backdoor-Vulnerabilities-in-Normally-Trained-Deep-Learning-Models, 2023-PELICAN}. 
Tao et al.~\cite{2022-Backdoor-Vulnerabilities-in-Normally-Trained-Deep-Learning-Models} show that normally trained deep learning models may contain latent backdoor-like vulnerabilities even without any malicious data poisoning, and that such vulnerabilities can be identified through trigger inversion. 
Furthermore, Zhang et al.~\cite{2023-PELICAN} demonstrate that naturally trained models in code-related scenarios can also exhibit exploitable backdoor-like behaviors, indicating that natural backdoors are not limited to traditional vision or natural language settings.
However, existing studies on natural backdoors in code-related settings remain limited. 
Therefore, this paper systematically investigates natural backdoors in \abbr{}s, with a particular focus on how such vulnerabilities emerge and how they manifest across different code intelligence tasks.}

\section{Conclusion}
\label{sec:conclusion}

In this paper, we conduct a systematic empirical study of backdoor vulnerabilities in \abbr{} trained on naturally occurring datasets across different models and code-related tasks. Our key findings are as follows: 1) Natural backdoors are widely present, and even large-scale \abbr{} cannot fully eliminate them; 2) Samples containing natural triggers exhibit highly covert behaviors in the representation space, remaining deeply entangled with clean samples and thus difficult to distinguish; 3) These vulnerabilities can transfer across \abbr{} fine-tuned on the same dataset, sharing the same architecture, or sharing learned knowledge; 4) Dataset bias is the primary cause of natural backdoors, whereas the training procedure itself plays a comparatively minor role; 5) \majorD{EliBadCode~\cite{2025-EliBadCode} can effectively mitigate such vulnerabilities through model unlearning.}\majorR{The unlearning-based defense can effectively mitigate such vulnerabilities.}
Furthermore, we propose a novel detection method to enhance the identification of natural backdoors in \abbr{}. Our findings highlight the severity of natural backdoor vulnerabilities and underscore the need for more effective defense techniques to strengthen the security of \abbr{}s.

\section{Data Availability}
Our source code and experimental data are available at~\cite{NatBackdoor}.

\bibliographystyle{IEEEtran}
\bibliography{reference}

\end{document}